\documentclass[11pt,a4paper]{article}
\usepackage{amsmath,amsthm,amsfonts,amssymb,mathrsfs}
\usepackage{graphicx}
\usepackage{subfigure}
\usepackage[latin1]{inputenc}
\usepackage[swedish,english]{babel}

\renewcommand{\vec}[1]{{\boldsymbol#1}}
\allowdisplaybreaks
\begin{document}
\title{Resonances in axially symmetric dielectric objects}
\author{Johan Helsing\thanks{Centre for Mathematical Sciences, Lund
    University, Sweden}~~and Anders Karlsson\thanks{Electrical and
    Information Technology, Lund University, Sweden}}
\date{\today}
\maketitle
\begin{abstract}
  A high-order convergent and robust numerical solver is constructed
  and used to find complex eigenwavenumbers and electromagnetic
  eigenfields of dielectric objects with axial symmetry. The solver is
  based on Fourier--Nyström discretization of combined integral
  equations for the transmission problem and can be applied to
  demanding resonance problems at microwave, terahertz, and optical
  wavelengths. High achievable accuracy, even at very high
  wavenumbers, makes the solver ideal for benchmarking and for
  assessing the performance of general purpose commercial software.
\end{abstract}

\section{Introduction}
\label{sec:intro}

This paper is about the fundamental problem of determining resonances
of axially symmetric homogeneous dielectric objects in vacuum. The
problem is formulated as an eigenvalue problem based on a combination
of the electric field integral equation (EFIE), the magnetic field
integral equation (MFIE), and two charge integral equations (ChIEs).
It is solved numerically using a high-order convergent discretization
scheme. A motivation for this work is to be able to produce very
accurate evaluations that can serve as bench mark tests for other
methods. This aim has lead us to use a formulation that, with our
scheme, gives the most accurate evaluations.

In microwave technology dielectric resonators are interesting since
they are cost effective and lead to significant miniaturization,
particularly of microwave integrated circuits. They give excellent
performance to antennas~\cite{Kishk02} and filters~\cite{Reaney06}. A
nice review of dielectric resonators in microwave technology is given
in~\cite{Fiedz02}. Resonant dielectric objects also play an important
role in the recent progress in nano-optics. A good example is axially
symmetric structures that exhibit whispering gallery modes
(WGMs)~\cite{Matsko06}. WGMs have large Q-factors (commercial
micro-optical WGM resonators can have ${\rm Q}>10^8$) and their
eigenfields are confined to a small volume in the outer part of the
dielectric object. These properties are very useful in the design of
microcavity lasers~\cite{He13} and extremely sensitive
sensors~\cite{Foreman15} and for the generation of frequency
combs~\cite{Chembo10,Kippenberg11}. WGMs have been used for
determining electric properties of materials~\cite{Rathe00} at
microwave frequencies, but otherwise WGMs have been less exploited in
microwave technology than in optics.

Common numerical methods for the determination of electromagnetic
resonances in dielectric objects include the finite element method
(FEM)~\cite{Oxborrow07}, boundary integral equation (BIE)
methods~\cite{Bulygin15,HelsKarl15,HelsKarl16}, the discrete dipole
approximation (DDA) method~\cite{Amendola10}, and the null-field
method~\cite{Zhen91}. The method used in the present paper belongs to
the category of BIE methods, which comprise a variety of formulations
and techniques. In \cite{Oijala05} the integral equations derived by
Müller~\cite[Section 23]{Muller69} was applied to scattering from
dielectric objects. We use a related set of integral equations and a
modification of the Fourier--Nyström scheme of~\cite{HelsKarl16},
which in turn draws on progress
in~\cite{HelsKarl15,Cohl99,HelsKarl14,HelsHols15,Youn12}.

Most BIE methods for transmission problems use the electric and
magnetic surface current densities as unknowns. A particular feature
of the present work is that we also let the surface charge densities
be unknowns. There are two reasons for this: First, the problem of
evaluating compact differences of hypersingular operators in the
classical Müller formulation is avoided. Second, numerical
differentiation of surface currents for the evaluation of eigenfields
is avoided. As a result, our scheme can be made higher-order and
attain extraordinary accuracy. It can easily solve resonance problems
that, to our knowledge, previously have been essentially inaccessible.

We remark that the use of unknown surface charge densities to improve
the performance of numerical schemes is not new. It was introduced as
a way to overcome the, so called, low frequency breakdown problem of
BIE methods in~\cite{TaskOija06} and was further developed for this
purpose in~\cite{Vico13}. See also \cite[Appendix A]{Gane14}.
In~\cite{HelsKarl15} and~\cite{HelsKarl16} it was recognized that the
use of unknown surface charges is numerically favorable also at higher
frequencies.

The paper is organized as follows: Section~\ref{sec:form} describes
the geometry and formulates our problem in terms of partial
differential equations (PDEs). Section~\ref{sec:intrepeq} introduces
integral representations of electric and magnetic fields in terms of
surface densities, derives the integral equations and discusses their
relation to the Müller integral equations. Section~\ref{sec:axialsym}
restricts the analysis to axially symmetric objects. Fourier series
expansions are used to express the homogeneous system of integral
equations, from which wavenumbers and surface densities representing
eigenfields are obtained, in a form that is well-suited for
discretization.  Section~\ref{sec:energy} defines useful physical
quantities and relate them to the surface densities. The numerical
method is described in Section \ref{sec:numerics}. Some challenging
numerical examples, involving various types of resonant modes, are
given in Section~\ref{sec:numex}.

\begin{figure}
  \centering \includegraphics[height=47mm]{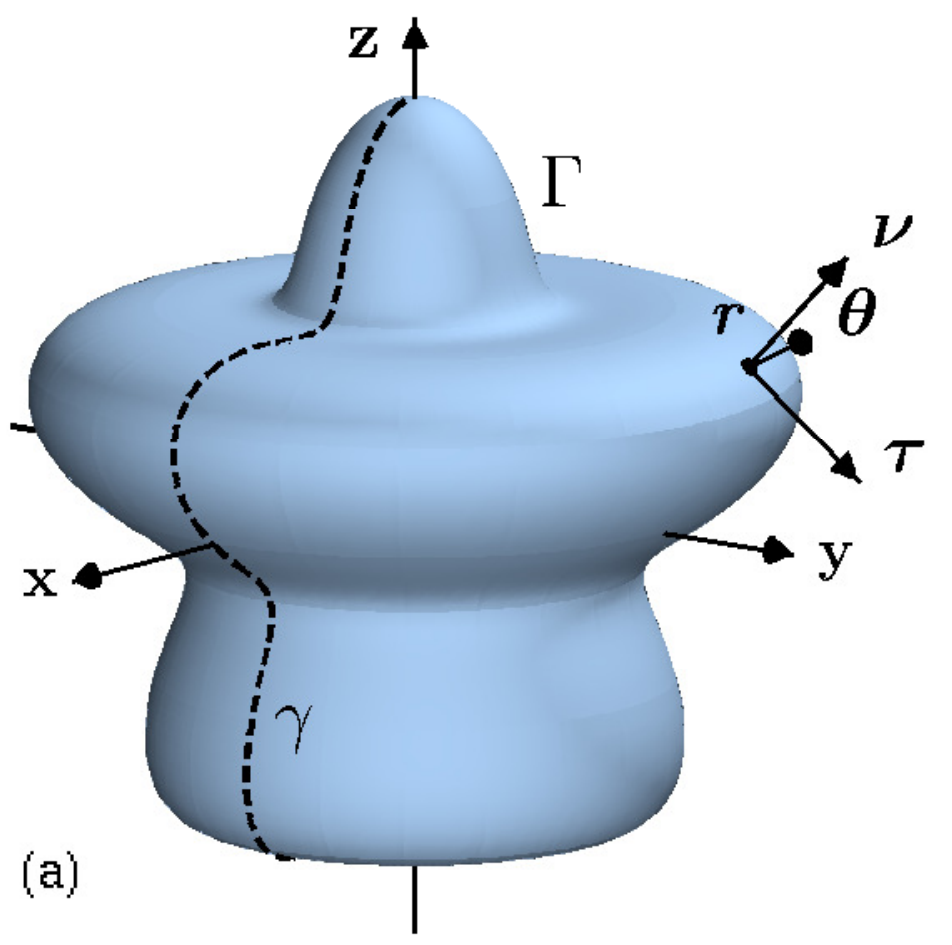}
  \includegraphics[height=47mm]{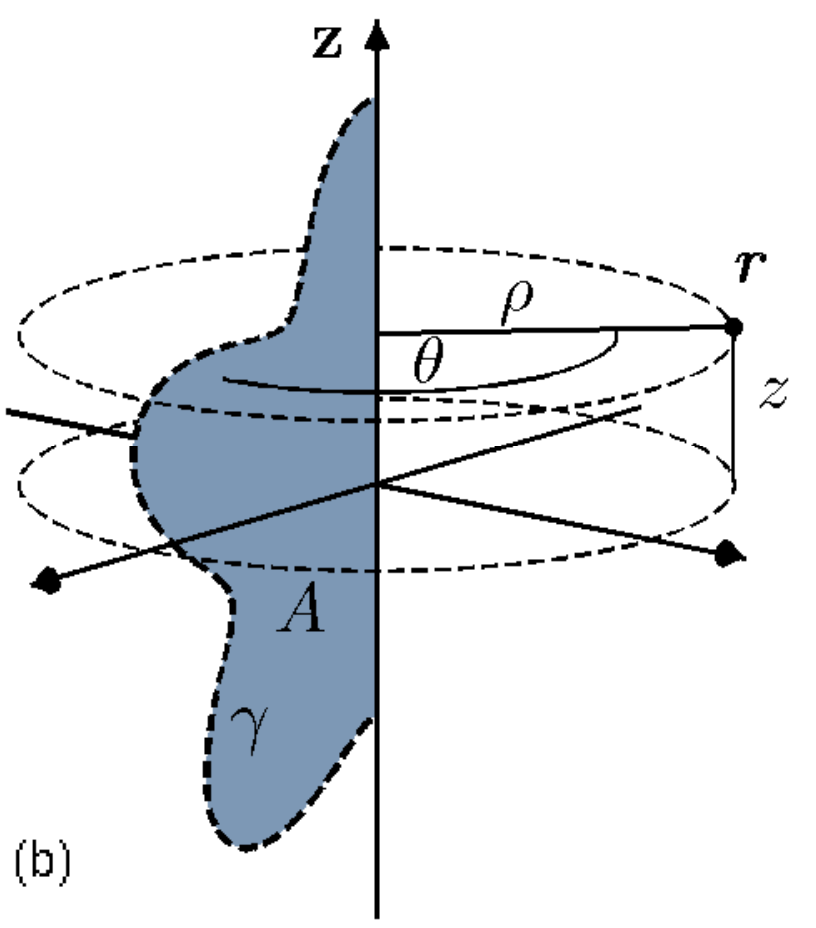}
  \includegraphics[height=47mm]{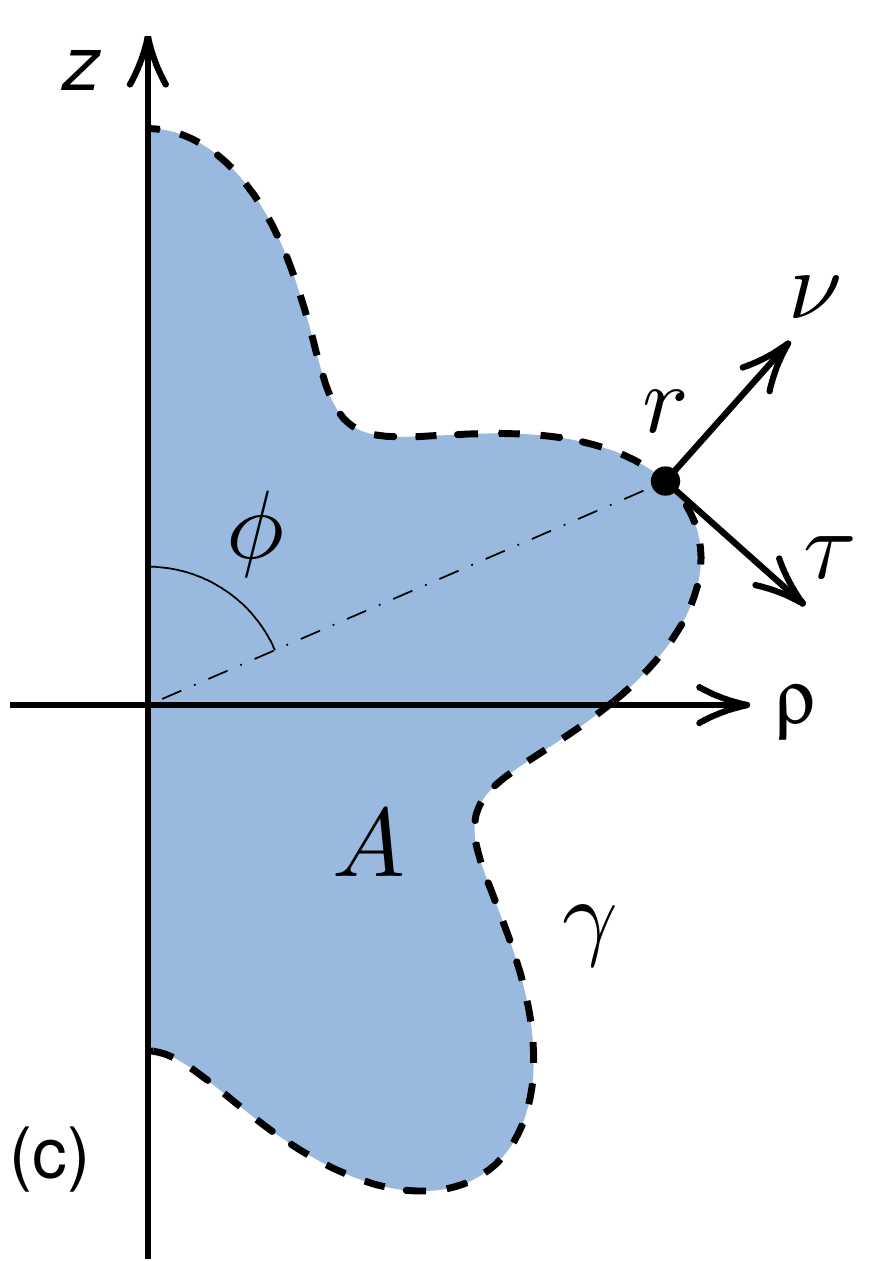}
\caption{\sf The geometry described in $\mathbb{R}^3$ and in
  $\mathbb{R}^2$. (a) The outward unit normal $\boldsymbol{\nu}$ and
  tangent vector $\boldsymbol{\tau}$ at a point ${\vec r}$ on
  $\Gamma$. The volume outside $\Gamma$ is $V_1$ and the volume inside
  is $V_2$. (b) The radial distance $\rho$, azimuthal angle $\theta$,
  and height $z$ of a point ${\vec r}$. The domain $A$ and the
  generating curve $\gamma$. (c) The half-plane $\mathbb{H}$ with
  two-dimensional vectors. }
\label{fig:geometry}
\end{figure}

\section{Problem formulation}
\label{sec:form}

\subsection{Geometry}
The notation is the same as in~\cite{HelsKarl16}. In particular,
$\Gamma$ is an axially symmetric surface enclosing a body of
revolution $V_2$ in $\mathbb{R}^3$, the unbounded exterior to
$\Gamma$ is $V_1$,
\begin{equation}
\vec r=\hat{\vec r}\vert\vec
r\vert=(x,y,z)=(\rho\cos\theta,\rho\sin\theta,z)
\end{equation}
is a point in $\mathbb{R}^3$, $\theta$ is the azimuthal angle,
$\rho=\sqrt{x^2+y^2}$, and $\hat{\vec r}$ is the radial unit vector.
The outward unit normal $\vec\nu$ on $\Gamma$ is
\begin{equation}
\vec\nu=(\nu_{\rho }\cos\theta,\nu_{\rho }\sin\theta,\nu_z)
\end{equation}
and
\begin{equation}
\begin{split}
\vec\rho&=(\cos\theta,\sin\theta,0)\,,\\
\vec\theta&=(-\sin\theta,\cos\theta,0)\,,\\
\vec\tau&=(\nu_z\cos\theta, \nu_z\sin\theta,-\nu_{\rho })\,,\\
\vec z&=(0,0,1)\,,
\end{split}
\end{equation}
are other unit vectors. See Figure~\ref{fig:geometry}(a)
and~\ref{fig:geometry}(b).

The angle $\theta=0$ defines a half-plane $\mathbb{H}$ in
$\mathbb{R}^3$. The intersection of $\mathbb{H}$ and $\Gamma$ is the
generating curve $\gamma$, points in $\mathbb{H}$ are denoted
$r=(\rho,z)$, the planar domain bounded by $\gamma$ and the $z$-axis
is $A$, the outward unit normal on $\gamma$ is $\nu=(\nu_\rho,\nu_z)$,
and $\tau=(\nu_z,-\nu_\rho)$ is a tangent. See
Figure~\ref{fig:geometry}(c).

\subsection{PDE-formulation}
The domain $V_2$ is a homogeneous dielectric object with constant
complex refractive index $m$. In $V_1$ there is vacuum. The electric
field is everywhere scaled with the free space wave impedance $\eta_0$
such that $\vec E=\eta_0^{-1}\vec E_{\rm unsc}$, where $\vec
E_{\rm unsc}$ is the unscaled field. Then $\vec E$ and the magnetic
field $\vec H$ have the same dimensions.

Sources can be located in a bounded volume $V_{{\rm s}1}$ in $V_1$ and
$V_{{\rm s}2}$ in $V_2$ and generate time harmonic incident fields
with complex electric and magnetic fields $\vec E^{\rm inc}_1$, $\vec
H^{\rm inc}_1$ in $V_1$ and $\vec E^{\rm inc}_2$, $\vec H^{\rm inc}_2$
in $V_2$. These give rise to the scattered fields $\vec E^{\rm sca}$
and $\vec H^{\rm sca}$ in $V_1$ and $V_2$. We prefer to work with the
total electric and magnetic fields $\vec E$ and $\vec H$, which are
the sum of the incident and scattered fields. From the Maxwell
equations it follows that the total fields satisfy the system of PDEs
\begin{align}
\nabla\times\vec E(\vec r)&={\rm i}k_j\vec H(\vec r)
\,,\quad\vec r\in V_j\setminus V_{{\rm s}j}\,,
\quad j=1,2\,,\label{eq:PDE1}\\
\nabla\times\vec H(\vec r)&=-{\rm i}k_j\vec E(\vec r)\,,
\quad\vec r\in V_j\setminus V_{{\rm s}j}\,,
\quad j=1,2\,,\label{eq:PDE2}\\
k_2&=mk_1\,,\label{eq:PDE4}
\end{align}
where $k_1=\omega/c$ and $k_2$ are the wavenumbers in $V_1$ and $V_2$,
$\omega$ is the angular frequency, and $c$ is the speed of light in
vacuum. We use the time dependence $e^{-{\rm i}\omega t}$. Then
$\Im{\rm m}\{m\}>0$, since the material in $V_2$ is assumed to be
passive. From now on we will, for the most part, omit the subscript of
$k_1$ and write the (vacuum) wavenumber in $V_1$ as $k$.

The boundary conditions on $\Gamma$ are
\begin{align}
\lim_{V_1\ni \vec r\to \vec r^\circ}\vec \nu^\circ\cdot\vec E(\vec r)&=
\lim_{V_2\ni \vec r\to \vec r^\circ}m^2\vec \nu^\circ\cdot\vec E(\vec r)\,, 
\,\vec r^\circ\in\Gamma,
\label{eq:BC1}\\
\lim_{V_1\ni \vec r\to \vec r^\circ}\vec\nu^\circ\times\vec E(\vec r)&=
\lim_{V_2\ni \vec r\to \vec r^\circ}\vec\nu^\circ\times\vec E(\vec r)\,, 
\,\vec r^\circ\in\Gamma,
\label{eq:BC2}\\
\lim_{V_1\ni \vec r\to \vec r^\circ}\vec\nu^\circ\times\vec H(\vec r)&=
\lim_{V_2\ni \vec r\to \vec r^\circ}\vec\nu^\circ\times\vec H(\vec r)\,, 
\,\vec r^\circ\in\Gamma,
\label{eq:BC3}\\
\lim_{V_1\ni \vec r\to \vec r^\circ}\vec \nu^\circ\cdot\vec H(\vec r)&=
\lim_{V_2\ni \vec r\to \vec r^\circ}\vec \nu^\circ\cdot\vec H(\vec r)\,,
\,\vec r^\circ\in\Gamma,
\label{eq:BC4}
\end{align}
and the radiation condition for the scattered field in $V_1$ is
\begin{equation}
\begin{split}
\vec E^{\rm sca}(\vec r)&=
\dfrac{e^{{\rm i}k\vert\vec r\vert}}{\vert\vec r\vert}
\left(\vec F(\hat{\vec r})+
\mathcal{O}\left(\dfrac{1}{\vert\vec r\vert}\right)\right), 
\, \vert\vec r\vert\rightarrow \infty,\\
\vec H^{\rm sca}(\vec r)&=
\dfrac{e^{{\rm i}k\vert\vec r\vert}}{\vert\vec r\vert}
\left(\hat{\vec r}\times\vec F(\hat{\vec r})+\mathcal{O}
\left(\dfrac{1}{\vert\vec r\vert}\right)\right), 
\,\vert\vec r\vert\rightarrow \infty,
\end{split}
\label{radiationcond}
\end{equation}
where $\vec F$ is the electric far-field pattern,
see~\cite[Eq.~(6.23)]{Colt13}.

Our resonance problem can now be formulated as follows: we seek
nontrivial solutions to \eqref{eq:PDE1}-\eqref{radiationcond} when the
incident fields are zero. The solutions are the eigenwavenumbers $k$
and the eigenfields $\vec E$ and $\vec H$.

\subsection{The radiation condition at complex wavenumbers}

We shall solve \eqref{eq:PDE1}-\eqref{radiationcond} using a BIE
method where the integral equations are derived from integral
representations of $\vec E$ and $\vec H$ containing the Green's
function
\begin{equation}
\Phi_k(\vec r,\vec r')=
\dfrac{e^{{\rm i}k\vert\vec r-\vec r'\vert}}{4\pi\vert\vec r-\vec r'\vert}\,.
\end{equation} 
The radiation condition \eqref{radiationcond} is then automatically
satisfied and says that, for real $k$ and in the far zone, the
scattered field is an outward traveling spherical vector wave.

At resonances, the eigenwavenumbers $k$ are complex with negative
imaginary part. The condition \eqref{radiationcond} then says that the
eigenfields grow exponentially at large distances. This is required
for the fields to satisfy causality and for the corresponding time
domain fields to be exponentially decaying as $e^{\Im{\rm m}\{k\}ct}$
in time. Causality says that the fields of a resonance at a time $t$
and at a distance $\vert\vec r\vert$ from an object left the object at
time $t-\vert\vec r\vert/c$. The attenuation implies that at time
$t-\vert\vec r\vert/c$, the fields in the object were $e^{-\Im{\rm
    m}\{k\}r}$ times stronger than at time $t$, in accordance with
\eqref{radiationcond}.

The condition \eqref{radiationcond} is vital in the derivation of
integral representations of $\vec E$ and $\vec H$ in $V_1$. In
\cite[Section IIC]{Wiersig02} it is shown that such derivations hold
also for eigenfields with complex eigenwavenumbers, despite their
exponential growth in the radial direction.

\section{Integral representations and equations}
\label{sec:intrepeq}

This section gives integral representations of $\vec E$ and $\vec H$
along with our system of BIEs for a general three-dimensional
dielectric object. The integral representations use four fictitious
surface densities on $\Gamma$: the magnetic and electric surface
current densities $\vec M_{\rm s}$ and $\vec J_{\rm s}$, and the
electric and magnetic surface charge densities $\varrho_{\rm E}$ and
$\varrho_{\rm M}$. The system of BIEs contains the EFIE, the MFIE, the
electric ChIE (EChIE), and the magnetic ChIE (MChIE).

\subsection{Surface densities and integral representations}
\label{sec:intrep}

The densities $\vec M_{\rm s}$, $\vec J_{\rm s}$, $\varrho_{\rm E}$,
and $\varrho_{\rm M}$ are defined from a viewpoint in $V_1$. With $\vec r^\circ\in\Gamma$:
\begin{align}
\hspace*{-2mm}\varrho_{\rm E}(\vec r^\circ)&\equiv
\lim_{V_1\ni \vec r\to \vec r^\circ}\vec \nu^\circ\cdot\vec E(\vec r)=
\lim_{V_2\ni \vec r\to \vec r^\circ}m^2\vec \nu^\circ\cdot\vec E(\vec r)
\label{eq:BC1def}\\
\hspace*{-2mm}\vec M_{\rm s}(\vec r^\circ)&\equiv
\lim_{V_1\ni \vec r\to \vec r^\circ}\vec E(\vec r)\times\vec \nu^\circ=
\lim_{V_2\ni \vec r\to \vec r^\circ}\vec E(\vec r)\times\vec \nu^\circ
\label{eq:BC2def}\\
\hspace*{-2mm}\vec J_{\rm s}(\vec r^\circ)&\equiv
\lim_{V_1\ni \vec r\to \vec r^\circ}\vec \nu^\circ\times\vec H(\vec r)=
\lim_{V_2\ni \vec r\to \vec r^\circ}\vec \nu^\circ\times\vec H(\vec r)
\label{eq:BC3def}\\
\hspace*{-2mm}\varrho_{\rm M}(\vec r^\circ)&\equiv
\lim_{V_1\ni \vec r\to \vec r^\circ}\vec \nu^\circ\cdot\vec H(\vec r)=
\lim_{V_2\ni \vec r\to \vec r^\circ}\vec \nu^\circ\cdot\vec H(\vec r)\,.
\label{eq:BC4def}
\end{align}
The second equalities in~(\ref{eq:BC1def})-(\ref{eq:BC4def}) hold when
the boundary conditions~(\ref{eq:BC1})-(\ref{eq:BC4}) are met.

Our integral representations of $\vec E(\vec r)$ and $\vec H(\vec r)$
are for $\vec r\in V_1$
\begin{equation}
\begin{split}
\vec E(\vec r)&=\vec E_1^{\rm inc}(\vec r)
-{\cal N}\varrho_{\rm E}(\vec r)
-{\cal K}\vec M_{\rm s}(\vec r)
+{\rm i}k{\cal S}\vec J_{\rm s}(\vec r)\,,\\
\vec H(\vec r)&=\vec H_1^{\rm inc}(\vec r)
+{\rm i}k{\cal S}\vec M_{\rm s}(\vec r)
+{\cal K}\vec J_{\rm s}(\vec r)
-{\cal N}\varrho_{\rm M}(\vec r)\,,
\end{split}
\label{eq:repr1}
\end{equation}
and for $\vec r\in V_2$
\begin{equation}
\begin{split}
\vec E(\vec r)&=\vec E_2^{\rm inc}(\vec r)
+m^{-2}\tilde{\cal N}\varrho_{\rm E}(\vec r)
+\tilde{\cal K}\vec M_{\rm s}(\vec r)
-{\rm i}k\tilde{\cal S}\vec J_{\rm s}(\vec r)\,,\\
\vec H(\vec r)&=\vec H_2^{\rm inc}(\vec r)
-{\rm i}m^2k\tilde{\cal S}\vec M_{\rm s}(\vec r)
-\tilde{\cal K}\vec J_{\rm s}(\vec r)
+\tilde{\cal N}\varrho_{\rm M}(\vec r)\,,
\end{split}
\label{eq:repr2}
\end{equation}
where the integral operators ${\cal S}$, ${\cal N}$, and ${\cal K}$
are defined by their actions on scalar or vector surface densities
$g(\vec r)$ and $\vec g(\vec r)$ as
\begin{align}
{\cal S}\vec g(\vec r)&=\int_\Gamma
\Phi_{k}(\vec r,\vec r')\vec g(\vec r')\,{\rm d}\Gamma'\,,
\label{eq:cS}\\
{\cal N}g(\vec r)&=\int_\Gamma
\nabla\Phi_{k}(\vec r,\vec r')g(\vec r')\,{\rm d}\Gamma'\,,
\label{eq:cN}\\
{\cal K}\vec g(\vec r)&=\int_{\Gamma}
\nabla\Phi_{k}(\vec r,\vec r')\times\vec g(\vec r')\,{\rm d}\Gamma'\,,
\label{eq:cK}
\end{align}
and $\tilde{\cal S}$, $\tilde{\cal N}$, and $\tilde{\cal K}$ are
defined analogously, but with $\Phi_k$ replaced by $\Phi_{k_2}$.

\subsection{Integral equations}
\label{sec:inteq}
We now form a system of BIE  on $\Gamma$. It comes from using~(\ref{eq:repr1})
and~(\ref{eq:repr2}) in the
definitions~(\ref{eq:BC1def})-(\ref{eq:BC4def}) and taking the limits
$\vec r\to\vec r^\circ\in\Gamma$. Each definition gives rise to two
BIEs: one for $\vec r\in V_1$ and one for $\vec r\in V_2$. 

The BIEs coming from $\vec r\in V_1$ are
\begin{equation}
\begin{split}
\varrho_{\rm E}+2\vec\nu\cdot\left({\cal N}\varrho_{\rm E}
+{\cal K}\vec M_{\rm s}-{\rm i}k{\cal S}\vec J_{\rm s}\right)&=
 2\vec\nu\cdot\vec E_1^{\rm inc}\,,\\
\vec M_{\rm s}-2\vec\nu\times\left({\cal N}\varrho_{\rm E}
+{\cal K}\vec M_{\rm s}-{\rm i}k{\cal S}\vec J_{\rm s}\right)&=
-2\vec\nu\times\vec E_1^{\rm inc}\,,\\
\vec J_{\rm s}-2\vec\nu\times\left({\rm i}k{\cal S}\vec M_{\rm s}
+{\cal K}\vec J_{\rm s}-{\cal N}\varrho_{\rm M}\right)&=
 2\vec\nu\times\vec H_1^{\rm inc}\,,\\
\varrho_{\rm M}-2\vec\nu\cdot\left({\rm i}k{\cal S}\vec M_{\rm s}
+{\cal K}\vec J_{\rm s}-{\cal N}\varrho_{\rm M}\right)&=
2\vec\nu\cdot \vec H_1^{\rm inc}\,.
\end{split}
\label{eq:syst1}
\end{equation}
The BIEs coming from $\vec r\in V_2$ are
\begin{equation}
\begin{split}
\varrho_{\rm E}-2\vec\nu\cdot\left(\tilde{\cal N}\varrho_{\rm E}
+m^2\tilde{\cal K}\vec M_{\rm s}
-{\rm i}m^2k\tilde{\cal S}\vec J_{\rm s}\right)&=
2m^2\vec\nu\cdot\vec E_2^{\rm inc}\,,\\
\vec M_{\rm s}+2\vec\nu\times\left(m^{-2}\tilde{\cal N}\varrho_{\rm E}
+\tilde{\cal K}\vec M_{\rm s}
-{\rm i}k\tilde{\cal S}\vec J_{\rm s}\right)&=
-2\vec\nu\times\vec E_2^{\rm inc}\,,\\
\vec J_{\rm s}+2\vec\nu\times\left({\rm i}m^2k\tilde{\cal S}\vec M_{\rm s}
+\tilde{\cal K}\vec J_{\rm s}-\tilde{\cal N}\varrho_{\rm M}\right)&=
2\vec\nu\times\vec H_2^{\rm inc}\,,\\
\varrho_{\rm M}+2\vec\nu\cdot\left({\rm i}m^2k\tilde{\cal S}\vec M_{\rm s} 
+\tilde{\cal K}\vec J_{\rm s}-\tilde{\cal N}\varrho_{\rm M}\right)&= 
2\vec\nu\cdot\vec H_2^{\rm inc}\,.
\end{split}
\label{eq:syst2}
\end{equation}
The order of the equations in~(\ref{eq:syst1}) and~(\ref{eq:syst2}) is
from top to bottom: EChIE, EFIE, MFIE, and MChIE.

We collect the systems~(\ref{eq:syst1}) and~(\ref{eq:syst2}) in block
operator form
 \begin{align}
{\bf Q}_1\vec\sigma&=\vec f_1\,,
\label{eq:Q1sys}\\
{\bf Q}_2\vec\sigma&=\vec f_2\,,
\label{eq:Q2sys}
\end{align}
where
\begin{equation}
\vec\sigma=
\begin{bmatrix}
\varrho_{\rm E}\\
\vec M_{\rm s}\\
\vec J_{\rm s}\\
\varrho_{\rm M}
\end{bmatrix},
\,\,\vec f_1=2
\begin{bmatrix}
\vec\nu\cdot\vec E_1^{\rm inc}\\-\vec\nu\times\vec E_1^{\rm inc}\\ 
\vec\nu\times\vec H_1^{\rm inc}\\ 
\vec\nu\cdot\vec H_1^{\rm inc}
\end{bmatrix},
\,\,\vec f_2=2
\begin{bmatrix}m^2\vec\nu\cdot\vec E_2^{\rm inc}\\ 
-\vec\nu\times\vec E_2^{\rm inc}\\ 
\vec\nu\times\vec H_2^{\rm inc}\\ 
\vec\nu\cdot\vec H_2^{\rm inc}
\end{bmatrix},
\label{eq:sigf}
\end{equation}
and where ${\bf Q}_1$ and ${\bf Q}_2$ are square block operator
matrices. The system \eqref{eq:Q1sys} is, modulo normalization
constants, identical to \cite[Eq.~18]{TaskOija06}. Our resonance
problem means that we must find simultaneous nontrivial solutions
to~(\ref{eq:Q1sys}) and~(\ref{eq:Q2sys}) when $\vec f_1=\vec f_2=\vec
0$.

\subsection{The ChIE-extended  formulation and its relation to the Müller BIE}
\label{subsec:chie}

We adopt a combination of \eqref{eq:Q1sys} and \eqref{eq:Q2sys} that
we refer to as the ChIE extended formulation since it contains the
electric and magnetic surface charge densities as unknowns.
\begin{equation}
({\bf Q}_1+{\bf Q}_2)\vec\sigma =\vec f_1+\vec f_2\,,
\label{eq:extended}
\end{equation}

In Section~\ref{sec:numex} we make a comparison between a scheme based
on \eqref{eq:extended} and a scheme based on the combination of BIE
presented by Müller in \cite[p. 319]{Muller69}. The Müller combination
is
\begin{equation}
{\bf Q}_3\vec\sigma_{\rm M} =\vec f_{\rm M}\,,
\label{eq:mullercomb1}
\end{equation}
where
\begin{equation}
\vec\sigma_{\rm M}=\begin{bmatrix}\vec M_{\rm s}\\\vec J_{\rm s}
\end{bmatrix}
\end{equation}
and
\begin{equation}
\vec f_{\rm M}=2
\begin{bmatrix}
-\vec\nu\times(\vec E_1^{\rm inc}+m^2\vec E_2^{\rm inc})\\
 \vec\nu\times(\vec H_1^{\rm inc}+\vec H_2^{\rm inc})
\end{bmatrix}.
\end{equation}
In accordance with~\cite{Oijala05}, we refer to it as the classical
Müller combination. It can be modified to a version that often is
preferred for method of moment schemes
\begin{equation}
{\bf Q}_4\vec\sigma_{\rm M} =\vec f_{\rm M}\,.
\label{eq:mullercomb2}
\end{equation}
The matrix operators in \eqref{eq:mullercomb1} and
\eqref{eq:mullercomb2} are 
\begin{equation}
{\bf Q}_3=\begin{bmatrix}(1+m^2)I-2\vec\nu\times({\cal K} 
-m^2\tilde{\cal K})& 
2{\rm i}k\vec\nu\times\left({\cal S}-m^2\tilde{\cal S}
+k^{-2}({\cal L}-\tilde{\cal L})\right)\\
-2{\rm i}k\vec\nu\times\left({\cal S}-m^2\tilde{\cal S}
+k^{-2}({\cal L}-\tilde{\cal L})\right)&
2I-2\vec\nu\times({\cal K}-\tilde{\cal K})
\end{bmatrix}
\label{QM}
\end{equation}
\begin{equation}
{\bf Q}_4=\begin{bmatrix}(1+m^2)I-2\vec\nu\times({\cal K}-m^2\tilde{\cal K})& 
2{\rm i}k\vec\nu\times\left({\cal S}-m^2\tilde{\cal S}
+k^{-2}({\cal P}-\tilde{\cal P})\right)\\
-2{\rm i}k\vec\nu\times\left({\cal S}-m^2\tilde{\cal S}
+k^{-2}({\cal P}-\tilde{\cal P})\right)&
2I-2\vec\nu\times({\cal K}-\tilde{\cal K})
\end{bmatrix}
\label{QMM}
\end{equation}
where
\begin{equation}\begin{split}
&{\cal L}\vec g(\vec r)=\nabla\int_\Gamma (\nabla\Phi_k(\vec r,\vec r'))\cdot\vec g(\vec r')\,{\rm d}\Gamma'\\
&{\cal P}\vec g(\vec r)=\int_\Gamma \nabla\Phi_k(\vec r,\vec r')\nabla'\cdot\vec g(\vec r')\,{\rm d}\Gamma'
\end{split}
\end{equation}
The combinations  \eqref{eq:mullercomb2}  and \eqref{eq:extended} are related via
\begin{align}
\varrho_{\rm E}(\vec r)&=
-\frac{\rm i}{k}\nabla_{\rm s}\cdot\vec J_{\rm s}(\vec r)\,,
\label{eq:checkE}\\
\varrho_{\rm M}(\vec r)&=
-\frac{\rm i}{k}\nabla_{\rm s}\cdot\vec M_{\rm s}(\vec r),
\label{eq:checkM}
\end{align}
where $\nabla_{\rm s}\cdot()$ is the surface divergence.

The operators ${\cal L}$ and $\tilde{\cal L}$ in \eqref{eq:mullercomb1}  are hypersingular, but  the
hypersingularities cancel out in the difference ${\cal L}-\tilde{\cal L}$. By that  \eqref{eq:mullercomb1} becomes a system
of Fredholm second kind integral equations with compact integral
operators~\cite[p.~300]{Muller69}. The hypersingularities are still present in the 
integral representations of the electric and magnetic fields and need to be handled by care in the evaluation of the fields close to $\Gamma$. It is  not straightforward to implement \eqref{eq:mullercomb1}. This is one reason why  \eqref{eq:mullercomb2}, which does not contain hypersingular operators, is an alternative. Our versions of ${\bf
  Q}_1$ and ${\bf Q}_2$ are free from hypersingular integral operators
but they do contain singular operators defined only in the sense of
the Cauchy principal value and do not all cancel out in the sum ${\bf
  Q}_1+{\bf Q}_2$. 

Müller showed two additional properties of his classical formulation
under the condition that $\Gamma$ consists of only one closed regular
surface:
\begin{enumerate}
\item The system~(\ref{eq:mullercomb1}) has a unique solution
  $\vec\sigma$ for wavenumbers $k$ with $0\leq{\rm arg}\{k\}<\pi$ and
  $0\leq{\rm arg}\{m^2\}<\pi/2$, \cite[Theorem 68]{Muller69}.
\item With $\vec\sigma$ as the unique solution
  to~(\ref{eq:mullercomb1}), the corresponding fields $\vec E$ and
  $\vec H$ obtained from \eqref{eq:repr1} and \eqref{eq:repr2} are
  solutions to the Maxwell equations \cite[Theorem 69]{Muller69}. By
  that they also satisfy \eqref{eq:PDE1}-\eqref{radiationcond}.
\end{enumerate}
A conjecture is that the ChIE-extended 
formulation~(\ref{eq:extended}) also has these properties.
Furthermore, in our numerical experiments with $\vec f_1=\vec f_2=\vec
0$ we check that all nontrivial solutions to~(\ref{eq:extended})
also solve \eqref{eq:Q1sys} and \eqref{eq:Q2sys} to the same precision
and that~(\ref{eq:checkE}) and~(\ref{eq:checkM}) hold (with a few
digits lost in the numerical differentiation). We have not been able
to detect any solution to~(\ref{eq:extended}) that violates these
tests.

\subsection{Physical resonances}

Our goal is to find eigenwavenumbers $k$ for the homogeneous version
of~(\ref{eq:extended})
\begin{equation}
({\bf Q}_1+{\bf Q}_2)\vec\sigma =\vec 0
\label{eq:homogen}
\end{equation}
and to evaluate their corresponding eigenfields $\vec E$ and $\vec H$
from \eqref{eq:repr1} and \eqref{eq:repr2} via $\vec\sigma$. The
eigenwavenumbers have $\Im{\rm m}\{k\}<0$ and constitute an infinite
countable set.

\section{Axial symmetry}
\label{sec:axialsym}

So far our analysis is valid for arbitrary dielectric objects. We now
restrict it to objects with axial symmetry and perform an azimuthal
Fourier transformation of (\ref{eq:Q1sys}) and (\ref{eq:Q2sys}) and of
(\ref{eq:repr1}) and (\ref{eq:repr2}) to obtain modal integral
equations and modal representations of $\vec E$ and $\vec H$. The
fields $\vec E$ and $\vec H$ are expressed in the cylindrical
coordinates $(\rho,\theta,z)$ as
\begin{equation}
\begin{split}
\vec E(\vec r)&=\vec\rho E_{\rho}(\vec r)
+\vec\theta E_{\theta}(\vec r)+\vec z E_{z}(\vec r)\,,\\
\vec H(\vec r)&=\vec\rho H_{\rho}(\vec r)
+\vec\theta H_{\theta}(\vec r)+\vec z H_{z}(\vec r)\,.
\end{split}
\end{equation}
The densities $\vec M_{\rm s}$ and $\vec J_{\rm s}$ are decomposed in
the two tangential directions $\vec \tau$ and $\vec \theta$, see
Figure \ref{fig:geometry}, as
\begin{equation}
\begin{split}
\vec M_{\rm s}(\vec r)&=\vec\tau M_\tau(\vec r)
+\vec\theta M_\theta(\vec r)\,,\\
\vec J_{\rm s}(\vec r)&=\vec\tau J_\tau(\vec r)
+\vec\theta J_\theta(\vec r)\,.
\end{split}
\label{eq:MJdecomp}
\end{equation}

\subsection{Fourier series expansions}
\label{sec:Fourier}

Let $g(\vec r)$ represent a surface density or a right hand side and
let $G$ represent an integral operator of Section~\ref{sec:intrep}
or~\ref{sec:inteq} with rotationally invariant kernel $G({\vec
  r},{\vec r}')$. The azimuthal Fourier coefficients $g_n(r)$ and
$G_n(r,r')$ of the functions $g(\vec r)$ and $G({\vec r},{\vec r}')$
are
\begin{align}
g_n(r)&=\frac{1}{\sqrt{2\pi}}\int_{-\pi}^{\pi}e^{-{\rm i}n\theta}
g({\vec r})\,{\rm d}\theta\,, 
\label{eq:gF}\\
G_n(r,r')&=
\frac{1}{\sqrt{2\pi}}\int_{-\pi}^{\pi}e^{-{\rm i}n(\theta-\theta')}
G({\vec r},{\vec r}')\,{\rm d}(\theta-\theta')\,.
\label{eq:GF}
\end{align}
The azimuthal index $n$ takes values $n=0,\pm 1,\pm 2,\ldots$. Modal
integral operators $G_n$ are defined in terms of the coefficients
$G_n(r,r')$ as
\begin{equation}
G_ng_n(r)=
\sqrt{2\pi}\int_\gamma G_n(r,r')g_n(r')\rho'\,{\rm d}\gamma'\,.
\label{eq:modalop}
\end{equation}
The singularities of $G(\vec r,\vec r')$ are inherited by $G_n(r,r')$
in the sense that weakly singular operators $G$ on $\Gamma$ correspond
to weakly singular operators $G_n$ on $\gamma$, and that the same
holds for Cauchy-type singular operators.

\subsection{Modal integral equations}

Using~(\ref{eq:MJdecomp}) and with the notation~(\ref{eq:gF}), the
Fourier coefficients of the vectors in~(\ref{eq:sigf}) each gets six
scalar entries (transformed scalar surface densities)

\begin{equation}
\vec\sigma_n=
\begin{bmatrix}
\varrho_{{\rm E}n}\\
 M_{\tau n}\\
 M_{\theta n}\\
 J_{\tau n}\\
 J_{\theta n}\\
 \varrho_{{\rm M}n}
\end{bmatrix},
\quad\vec f_{1n}=2
\begin{bmatrix}
 E^{\rm inc}_{1\nu n}\\
 E^{\rm inc}_{1\theta n}\\
-E^{\rm inc}_{1\tau n}\\
-H^{\rm inc}_{1\theta n}\\
 H^{\rm inc}_{1\tau n}\\
 H^{\rm inc}_{1\nu n}\\
\end{bmatrix}, 
\quad\vec f_{2n}=2
\begin{bmatrix}
 m^2E^{\rm inc}_{2\nu n}\\
 E^{\rm inc}_{2\theta n}\\
-E^{\rm inc}_{2\tau n}\\
-H^{\rm inc}_{2\theta n}\\
 H^{\rm inc}_{2\tau n}\\
 H^{\rm inc}_{2\nu n}
\end{bmatrix}.
\end{equation}

The modal counterpart of~(\ref{eq:Q1sys}) and~(\ref{eq:Q2sys}) becomes
\begin{align}
{\bf Q}_{1n}\vec\sigma_n&=\vec f_{1n}\,,
\label{eq:Q1sysF}\\
{\bf Q}_{2n}\vec\sigma_n&=\vec f_{2n}\,.
\label{eq:Q2sysF}
\end{align}
The block operator matrices ${\bf Q}_{1n}$ and ${\bf Q}_{2n}$ can be
written
\begin{equation}
{\bf Q}_{1n}=
\begin{bmatrix}
 I+2K_{\nu n}&-2{\rm i}K_{25 n}&2K_{26 n}&-2{\rm i}kS_{5n}&2kS_{6 n}&0\\
-2{\rm i}K_{12 n}&I-K_{1n}&-{\rm i}K_{2n}&2kS_{3n}&-2{\rm i}kS_{4n}&0\\
 2K_{24 n}&-{\rm i}K_{3n}&I-K_{4n}&2{\rm i}kS_{1n}&-2kS_{2n}&0\\
 0&-2kS_{3n}&2{\rm i}kS_{4n}&I-K_{1n}&-{\rm i}K_{2n}&2{\rm i}K_{12n}\\
 0&-2{\rm i}kS_{1n}&2kS_{2n}&-{\rm i}K_{3n}&I-K_{4n}&-2K_{24n}\\
 0&-2{\rm i}kS_{5n}&2kS_{6n}&2{\rm i}K_{25n}&-2K_{26n}&I+2K_{\nu n}
\end{bmatrix}
\label{eq:Q1n}
\end{equation}
and
\begin{equation}
{\bf Q}_{2n}=
\begin{bmatrix}
 I-2\tilde K_{\nu n}&2{\rm i}m^2\tilde K_{25n}&-2m^2\tilde K_{26n}&
 2{\rm i}m^2k\tilde S_{5n}&-2m^2k\tilde S_{6n}&0\\
 2{\rm i}m^{-2}\tilde K_{12n}&I+\tilde K_{1n}&{\rm i}\tilde K_{2n}&
-2k\tilde S_{3n}&2{\rm i}k\tilde S_{4n}&0\\
-2m^{-2}\tilde K_{24n}&{\rm i}\tilde K_{3n}&I+\tilde K_{4n}&
-2{\rm i}k\tilde S_{1n}&2k\tilde S_{2n}&0\\
 0&2m^2k\tilde S_{3n}&-2{\rm i}m^2k\tilde S_{4n}&I+\tilde K_{1n}&
 {\rm i}\tilde K_{2n}&-2{\rm i}\tilde K_{12n}\\
 0&2{\rm i}m^2k\tilde S_{1n}&-2m^2k\tilde S_{2n}&{\rm i}\tilde K_{3n}&
 I+\tilde K_{4n}&2\tilde K_{24n}\\
 0&2{\rm i}m^2k\tilde S_{5n}&-2m^2k\tilde S_{6n}&
-2{\rm i}\tilde K_{25 n}&2\tilde K_{26n}&I-2\tilde K_{\nu n}
\end{bmatrix},
\label{eq:Q2n}
\end{equation}
where $I$ is the identity, $S_{in}$ and $K_{in}$, with various indices
$i$, are modal operators stemming from ${\cal S}$, ${\cal N}$, and
${\cal K}$ and generally defined via~(\ref{eq:GF})
and~(\ref{eq:modalop}), and the tilde symbol means the replacement of
$k$ by $k_2$ as explained in Section~\ref{sec:intrep}. The operators
$S_{in}$ are weakly singular. The $K_{in}$ are weakly singular for
$i=\nu,1,2,3,4$ and Cauchy-type singular for $i=12,24,25,26$.

All modal operators in~(\ref{eq:Q1n}) and~(\ref{eq:Q2n}) are detailed
in~\cite[Appendix~A]{HelsKarl16} except for $K_{in}$, $i=24,25,26$,
which are given in Appendix~A of the present paper.

\subsection{Modal representations of $\vec E$ and $\vec H$}
\label{sec:EHF}

Once the modal counterpart of~(\ref{eq:extended}),
\begin{equation}
\left({\bf Q}_{1n}+{\bf Q}_{2n}\right)\vec\sigma_n=\vec f_{1n}+\vec f_{2n}\,,
\label{eq:extendedF}
\end{equation}
has been solved for $\vec\sigma_n$, modal representations of $\vec E$
and $\vec H$ can be constructed from modal counterparts of
\eqref{eq:repr1} and~\eqref{eq:repr2}.

The modal representations of the fields in $V_1$ are
\begin{equation}
\begin{split}
E_{\rho n}(r)&=K_{11n}\varrho_{{\rm E}n}-{\rm i}K_{5n}M_{\tau n}
-K_{6n}M_{\theta n}+{\rm i}k S_{7n}J_{\tau n}+kS_{8n}J_{\theta n}\,,\\
E_{\theta n}(r)&={\rm i}K_{12n}\varrho_{{\rm E}n}-K_{7n}M_{\tau n}
-{\rm i}K_{8n}M_{\theta n}+kS_{9n} J_{\tau n}+{\rm i}kS_{10n}J_{\theta n}\,,\\
E_{zn}(r)&=K_{13n}\varrho_{{\rm E}n}-{\rm i}K_{9n}M_{\tau n}
-K_{10n}M_{\theta n}+{\rm i}kS_{11n}J_{\tau n}\,,\\
\end{split}
\label{eq:eigenE1}
\end{equation}
and
\begin{equation}
\begin{split}
H_{\rho n}(r)&={\rm i}k S_{7n}M_{\tau n}+kS_{8n}M_{\theta n}
+{\rm i}K_{5n}J_{\tau n}+K_{6n}J_{\theta n}+K_{11n}\varrho_{{\rm M}n}\,,\\
H_{\theta n}(r)&=kS_{9n} M_{\tau n}+{\rm i}kS_{10n}M_{\theta n}
+K_{7n}J_{\tau n}+{\rm i}K_{8n}J_{\theta n}
+{\rm i}K_{12n}\varrho_{{\rm M}n}\,,\\
H_{zn}(r)&={\rm i}kS_{11n}M_{\tau n}+{\rm i}K_{9n}J_{\tau n}
+K_{10n}J_{\theta n}+K_{13n}\varrho_{{\rm M}n}\,.
\end{split}
\label{eq:eigenH1}
\end{equation}
The modal representations of the fields in $V_2$ are
\begin{equation}
\begin{split}
E_{\rho n}(r)&=-m^{-2}\tilde{K}_{11n}\varrho_{{\rm E}n}
+{\rm i}\tilde{K}_{5n}M_{\tau n}+\tilde{K}_{6n}M_{\theta n}
-{\rm i}k \tilde{S}_{7n}J_{\tau n}-k\tilde{S}_{8n}J_{\theta n}\,,\\
E_{\theta n}(r)&=-{\rm i}m^{-2}\tilde{K}_{12n}\varrho_{{\rm E}n}
+\tilde{K}_{7n}M_{\tau n}+{\rm i}\tilde{K}_{8n}M_{\theta n}
-k\tilde{S}_{9n}J_{\tau n}-{\rm i}k\tilde{S}_{10n}J_{\theta n}\,,\\
E_{zn}(r)&=-m^{-2}\tilde{K}_{13n}\varrho_{{\rm E}n}
+{\rm i}\tilde{K}_{9n}M_{\tau n}+\tilde{K}_{10n}M_{\theta n}
-{\rm i}k\tilde{S}_{11n}J_{\tau n}\,,
\end{split}
\label{eq:eigenE2}
\end{equation}
and
\begin{equation}\begin{split}
H_{\rho n}(r)&=-{\rm i}m^2k \tilde{S}_{7n}M_{\tau n}
-m^2k\tilde{S}_{8n}M_{\theta n}-{\rm i}\tilde{K}_{5n}J_{\tau n}
-\tilde{K}_{6n}J_{\theta n}-\tilde{K}_{11n}\varrho_{{\rm M}n}\,,\\
H_{\theta n}(r)&=-m^2k\tilde{S}_{9n} M_{\tau n}
-{\rm i}m^2k\tilde{S}_{10n}M_{\theta n}-\tilde{K}_{7n}J_{\tau n}
-{\rm i}\tilde{K}_{8n}J_{\theta n}
-{\rm i}\tilde{K}_{12n}\varrho_{{\rm M}n}\,,\\
H_{zn}(r)&=-{\rm i}m^2k\tilde{S}_{11n}M_{\tau n}
-{\rm i}\tilde{K}_{9n}J_{\tau n}-\tilde{K}_{10n}J_{\theta n}
-\tilde{K}_{13n}\varrho_{{\rm M}n}\,.
\end{split}
\label{eq:eigenH2}
\end{equation}
Here the operators $S_{in}$ and $K_{in}$, with various indices $i$,
are detailed in~\cite{HelsKarl15} and~\cite[Appendix~A]{HelsKarl16}.

\subsection{Eigenwavenumbers, eigenfields, and fundamental modes}
\label{sec:eigen}

The eigenwavenumbers at a prescribed refractive index $m$ are
wavenumbers $k$ for which, for some azimuthal index $n$, there exist
nontrivial solutions $\vec\sigma_n$ to the homogeneous version
of~(\ref{eq:extendedF})
\begin{equation}
\left({\bf Q}_{1n}+{\bf Q}_{2n}\right)\vec\sigma_n=\vec 0\,.
\label{eq:eigenQF}
\end{equation}
Our resonance problem now means finding such numbers $k$,
corresponding eigendensities $\vec\sigma_n$, and eigenfields $\vec
E_n(\vec r)$, $\vec H_n(\vec r)$ represented
by~(\ref{eq:eigenE1})-(\ref{eq:eigenH2}).

In \cite{HelsKarl16} it was shown how to form the physical time-domain
fields $\vec E_n(\vec r,t)$ from the Fourier coefficient vector $\vec
E_n(r)$. It was shown that one can let $E_{\rho n}(r)=E_{\rho
  (-n)}(r)$. Then the physical component $E_{\rho n}(\vec r,t)$
becomes
\begin{equation}
E_{\rho n}(\vec r,t)=
\dfrac{1}{2}\Re{\rm e}\{(E_{\rho n}(r)e^{{\rm i}n\theta}
+E_{\rho (-n)}(r)e^{-{\rm i}n\theta})e^{{\rm i}\omega t}\}\,.
\end{equation}
Since $\omega=kc$, we get
\begin{equation}
\omega
\equiv\omega_{\rm r}-{\rm i}\alpha
=\Re{\rm e}\{k\}c+{\rm i}\Im{\rm m}\{k\}c\,,
\label{eq:omegar}
\end{equation}
where $\omega_{\rm r}$ is the real angular frequency and $\alpha$ is
the attenuation constant. Then
\begin{equation}
E_{\rho n}(\vec r,t)=\Re{\rm e}\{E_{\rho n}(r)e^{-{\rm i}\omega_{\rm r} t}\} 
\cos n\theta e^{-\alpha t}\,.
\label{eq:physeig}
\end{equation}
This is a standing wave in the azimuthal direction. Also $E_{zn}(\vec
r,t)$ and $H_{\theta n}(\vec r,t)$ are proportional to $\cos n\theta$
whereas $E_{\theta n}(\vec r,t)$, $H_{\rho n}(\vec r,t)$, and
$H_{zn}(\vec r,t)$ are proportional to $\sin n\theta$. If one lets
$E_{\rho n}(r)=-E_{\rho (-n)}(r)$, then $\cos n\theta$ and $\sin
n\theta$ are exchanged in all components.

It is convenient to introduce the concept of the fundamental mode. The
fundamental mode, for a given $n$, is the resonance with the smallest
value of $\vert\Re{\rm e}\{k\}\vert$. For large $n$, it has properties
that distinguishes it from other resonances: Its electric and magnetic
fields are confined to a small volume in $V_2$ and are strongly
attenuated in the proximity of that small volume. The exponential
growth of the fields, in concordance with \eqref{radiationcond}, is
only seen at large distances. The ratio $\omega_{\rm r}/\alpha$ is
large. Fundamental modes with large $n$ are whispering gallery modes and are important in
optical applications as described in Section~\ref{sec:intro}.

\section{Powers, energies, and far-fields}
\label{sec:energy}
There is
no inner product under which the eigenfields are orthogonal and it is
also impossible to uniquely define a stored energy, a radiated power
of an eigenfield, and a normalization. It is, nevertheless, relevant
to introduce approximate expressions for these quantities and to
define the related Q-factor. We define the stored energy as the
electromagnetic energy stored in $V_2$ and define the radiated power
as the power radiated from $\Gamma$. The sharpness, or quality, of
these definitions becomes better as $\vert\Re{\rm
  e}\{k\}\vert/\vert\Im{\rm m}\{k\}\vert$ increases.

Assume a single resonance with azimuthal index $n$ that is excited by
an incident field for $t<0$ and that there are no incident fields for
$t\geq 0$. According to \eqref{eq:physeig}, the physical eigenfield
then oscillates with the angular frequency $\omega_{\rm r}$ and
attenuates as $e^{-\alpha t}$ for $t\geq0$. For $t\geq 0$ we let
$P_{\rm rad}(t)$ denote the radiated power through $\Gamma$ averaged
over one period $[t,t+T]$, $P_{\rm diss}(t)$ the dissipated power
averaged over the same period, and $W(t)$ the stored electromagnetic
energy in $V_2$ at time $t$. Conservation of energy and
\eqref{eq:physeig} lead to the relations
\begin{align}
P(t)&=P_{\rm rad}(t)+P_{\rm diss}(t)\,,
\label{eq:psum}\\
P_{\rm rad}(t)&=P_{\rm rad}(0)e^{-2\alpha t}\,,\\
P_{\rm diss}(t)&=P_{\rm diss}(0)e^{-2\alpha t}\,,
\label{eq:qnergyrel1}\\
W(t)&=W(0)e^{-2\alpha t}\,,
\label{eq:qnergyrel2}\\
W(0)&\equiv\int_0^\infty P(t)\,{\rm d}t
=\dfrac{P_{\rm rad}(0)+P_{\rm diss}(0)}{2\alpha}\,.
\label{eq:qnergyrel3}
\end{align}

We use the standard definition of the Q-factor which,
with~(\ref{eq:omegar}), can be written
\begin{equation}
{\rm Q}\equiv\omega_{\rm r}\dfrac{W(t)}{P(t)}
=-\dfrac{\Re{\rm e}\{k\}}{2\Im{\rm m}\{k\}}\,,
\label{eq:Q}
\end{equation}
and introduce
\begin{align}
{\rm Q}_{\rm rad}&=\omega_{\rm r}\dfrac{W(t)}{P_{\rm rad}(t)}\,,\\
{\rm Q}_{\rm diss}&=\omega_{\rm r}\dfrac{W(t)}{P_{\rm diss}(t)}\,,
\end{align}
so that, from \eqref{eq:psum} and \eqref{eq:Q},
\begin{equation}
\dfrac{1}{{\rm Q}}=
\dfrac{1}{{\rm Q}_{\rm rad}}+\dfrac{1}{{\rm Q}_{\rm diss}}\,.
\label{eq:Qfactors}
\end{equation}
The radiated power from $V_2$ equals the real part of the Poynting
vector integrated over $\Gamma$
\begin{equation}
\begin{split}
P_{\rm rad}(t)&=\dfrac{1}{2}\Re{\rm e}\left\{\int_\gamma\nu\cdot(\vec E_n(r)
\times\vec H^*_n(r))\rho\,{\rm d}\gamma\right\}e^{-2\alpha t}\\
 &=\dfrac{1}{2}\Re{\rm e}\left\{\int_\gamma( M_{\theta n}(r) J^*_{\tau n}(r)
-M_{\tau n}(r)J^*_{\theta n}(r))\rho\,{\rm d}\gamma\right\}e^{-2\alpha t}\,.
\end{split}
\label{eq:P}
\end{equation}
From Gauss theorem and the Maxwell equations it also follows that
\begin{equation}
P_{\rm rad}(0)=
-P_{\rm diss}(0)-\Im{\rm m}\{k\}\int_A\left(\vert \vec H_n(r)\vert^2
+\Re{\rm e}\left\{m^2\right\}\vert\vec E_n(r)\vert^2\right)\rho\,{\rm d}A\,,
\label{eq:P2}
\end{equation}
and due to \eqref{eq:qnergyrel3},
\begin{equation}
W(0)=\dfrac{1}{2c}\int_A\left(\vert\vec H_n(r)\vert^2
+\Re{\rm e}\left\{m^2\right\}\vert\vec E_n(r)\vert^2\right)\rho\,{\rm d}A\,.
\label{eq:energy0}
\end{equation}
Apart from a scale factor $\eta_0$, see Section~\ref{sec:form}, the
expression~(\ref{eq:energy0}) is the standard expression for the
electromagnetic energy in a volume.

The skin depth
\begin{equation}
\delta=(\Im{\rm m}\{m\}\Re{\rm e}\{k\})^{-1}
\end{equation}
is a measure of $\Im{\rm m}\{m\}$. It is derived from the attenuation
of a plane wave that impinges at normal incidence on a lossy half
space, but is also a measure of the attenuation of waves in dielectric
objects. When
\begin{equation}
\delta\gg {\rm diam}(V_2)\,, 
\label{eq:skindepth} 
\end{equation} 
a number of approximations are valid: The electric and magnetic
eigenfields inside $V_2$ and on $\Gamma$ are, to a high degree,
independent of $\Im{\rm m}\{m\}$. It then follows, from (\ref{eq:P})
and (\ref{eq:energy0}), that ${\rm Q}_{\rm rad}$ is independent of
$\Im{\rm m}\{m\}$. It also holds that
\begin{equation}
{\rm Q}_{\rm diss}\approx\dfrac{\Re{\rm e}\{m\}}{2\Im{\rm m}\{m\}}\,.
\label{eq:Qapp} 
\end{equation}

The normalized far-field pattern of a mode with azimuthal index $n$ is the $\phi-$dependent function
\begin{equation}
\dfrac{\vert\vec F_n(\phi)\vert}
{\max\limits_{0\leq\phi\leq\pi}\vert \vec F_{n}(\phi)\vert}\,,
\label{eq:farfield1}
\end{equation}
where $\vec F_n(\phi)$ is the Fourier coefficient of $\vec F$ in
\eqref{radiationcond}, and by that
\begin{equation}
\vec F_n(\phi)=\lim\limits_{\vert r\vert\rightarrow\infty}
e^{-{\rm i}k\vert r\vert}\vert r\vert\vec E_n(r)\,.
\label{eq:farfield2}
\end{equation}
The far-field pattern tells us in what directions the stored energy in
$V_2$ is radiated. By reciprocity it also indicates what direction an
incident wave should have in order to excite a resonance. Far-field
patterns are included in the numerical examples of
Section~\ref{sec:numex}.

\section{Discretization}
\label{sec:numerics}

The Fourier--Nyström discretization scheme for~(\ref{eq:eigenQF}) is adopted from~\cite{HelsKarl16}. This section gives a
brief overview and describes some modifications that are appropriate
when solving~(\ref{eq:eigenQF}) at high wavenumbers.

\subsection{Overview}
\label{sec:over}

Let $G_n$ be a generic modal integral operator of the type encountered
in~(\ref{eq:eigenQF}) and let $G_n(r,r')$ and $G(\vec r,\vec r')$ be
related to $G_n$ as in Section~\ref{sec:Fourier}. We split $G_n(r,r')$
into a smooth and a non-smooth function
\begin{equation}
G_n(r,r')=G_n^{(\rm s)}(r,r')+G_n^{(\rm ns)}(r,r')\,,
\label{eq:ksplit}
\end{equation}
where $G_n^{(\rm s)}(r,r')$ is zero when $r$ and $r'$ lie close to
each other and $G_n^{(\rm ns)}(r,r')$ is zero otherwise. We also split
$G(\vec r,\vec r')$ analogously. The kernel split~(\ref{eq:ksplit})
corresponds to an operator split $G_n=G_n^{(\rm s)}+G_n^{(\rm ns)}$.

The discretization of a $G_n$ in~(\ref{eq:modalop}) results in a
square matrix whose entries are values of $G_n(r,r')$, obtained from
$G(\vec r,\vec r')$ via~(\ref{eq:GF}), multiplied with suitable
quadrature weights. As underlying quadrature rules we use the
trapezoidal rule in~(\ref{eq:GF}) and 16th-order panel-based
Gauss--Legendre quadrature in~(\ref{eq:modalop}). This is sufficient
for the accurate discretization of the $G_n^{(\rm s)}$.

The efficient discretization of a $G_n^{(\rm ns)}$ requires that a
number of techniques are activated, all of which are described in
detail in~\cite{HelsKarl15,HelsKarl16,HelsKarl14}. The most important
are: evaluation of the integral over $G^{(\rm ns)}(\vec r,\vec r')$
in~(\ref{eq:GF}) via factorization and convolution; use of fast
discrete Fourier transform techniques and half-integer degree Legendre
functions of the second kind~\cite{Grad07}
\begin{equation}
\mathfrak{Q}_{n-\frac{1}{2}}(\chi)=
\int_{-\pi}^{\pi}\frac{\cos(nt)\,{\rm d}t}
{\sqrt{8\left(\chi-\cos(t)\right)}}\,,
\end{equation}
with
\begin{equation}
\chi=1+\frac{|r-r'|^2}{2\rho\rho'}\,,
\label{eq:Qndef}
\end{equation}
to evaluate the Fourier coefficients needed in this convolution;
16th-order accurate product integration for singular integrals on
$\gamma$, constructed on-the-fly and based on known asymptotics of
$\mathfrak{Q}_{n-\frac{1}{2}}(\chi)$ as $\chi\to 1^+$; a strategy for
when to use forward or backward recursion for the evaluation of
$\mathfrak{Q}_{n-\frac{1}{2}}(\chi)$; temporary mesh refinement
(upsampling) coupled with temporary increase of the quadrature order
on $\gamma$.

The discretization of the modal representation of $\vec E$ and $\vec
H$ for $r\notin\gamma$ in Section~\ref{sec:EHF} is done in analogy
with the discretization of~(\ref{eq:eigenQF}).

\subsection{Modifications}

The discretization of a $G_n^{(\rm ns)}$ becomes more difficult as $n$
grows. The domain where the known asymptotics of its kernel is useful
becomes narrower and forward recursion for
$\mathfrak{Q}_{n-\frac{1}{2}}(\chi)$ becomes increasingly unstable. In
previous work we let each quadrature panel along $\gamma$ be
temporarily divided into at most four subpanels for the resolution of
$\mathfrak{Q}_{n-\frac{1}{2}}(\chi)$ at arguments close to unity and
we used (expensive but stable) backward recursion whenever
$\chi>1.0005$. Here we allow up to six subpanels and use backward
recursion, as in~\cite{Gil07}, whenever $\chi>1.0001$.

Eigenwavenumbers are found with Broyden's method, which is one of the
simplest and most effective secant updating method for solving
nonlinear systems~\cite{Heath02}: Let $\lambda(k)$ be the smallest
magnitude eigenvalue of the system matrix in~(\ref{eq:eigenQF}) at
wavenumber $k$. We seek eigenwavenumbers $k$ as solutions to the
system
\begin{equation}
\begin{split}
\Re{\rm e}\left\{\lambda(k)\right\}&=0\,,\\
\Im{\rm m}\left\{\lambda(k)\right\}&=0\,,
\end{split}
\end{equation}
where $\Re{\rm e}\{k\}$ and $\Im{\rm m}\{k\}$ are considered
independent unknowns. For an initial guess $k$ that is reasonably
close to a zero of $\lambda(k)$, Broyden's method converges to almost
full achievable precision in about ten iterations.

Whispering gallery modes (WGMs) with high indices $n$ have eigendensities $\vec\sigma_n$ with
numerically discernible support only on those parts of $\gamma$ that
lie farthest away from the $z$-axis. We exploit this property to
reduce the number of unknowns when discretizing~(\ref{eq:eigenQF}) in
the search for high-index WGMs.

\section{Numerical examples}
\label{sec:numex}

We have implemented our Fourier--Nyström scheme for~(\ref{eq:eigenQF})
and~(\ref{eq:eigenE1})-(\ref{eq:eigenH2}) in {\sc Matlab}, release
2014a. We use a standard implementation and built-in functions. Our
workstation has 64 GB of memory and an Intel Core i7-3930K CPU.

The examples we are about to present share some common features:
\begin{itemize}
\item The dielectric object is either the unit sphere or the object in
  Figure~\ref{fig:geometry} whose generating curve $\gamma$ has the
  parameterization
\begin{equation}
r(s)=(1+0.25\cos(5s))(\sin(s),\cos(s))\,,\qquad 0\leq s\leq\pi\,.
\label{eq:star}
\end{equation}
\item The refractive index is either $m=1.5$ or $m=1.5+5.5\cdot
  10^{-12}{\rm i}$.
\item The planar field plots show the absolute values of some of the
  coefficients $(H_{\rho n}(r),H_{\theta n}(r),H_{zn}(r))$ and
  $(E_{\rho n}(r),E_{\theta n}(r),E_{zn}(r))$. In each example, all
  six coefficients are evaluated and scaled with a common factor so
  that the largest pointwise value of at least one coefficient is
  unity. The coefficients are evaluated at $5\cdot 10^5$ points $r$ on
  a Cartesian grid in a rectangle of height 2.6 and width 1.3 and with
  its left side coinciding with the $z$-axis in the half-plane
  depicted in Figure~\ref{fig:geometry}(c). For ease of
  interpretation, we also show mirror images so that a field plot
  includes $10^6$ points in a square of side length 2.6 in the
  $xz$-plane.
\item The estimated errors in the field plots are taken as the
  absolute value of the pointwise difference to a reference solution.
  In the absence of semi-analytic solutions, the reference solution is
  obtained with an overresolved mesh containing 50 per cent more
  quadrature panels on $\gamma$. The error plots use a logarithmic
  scale.
\end{itemize}
Our examples cover two modal cases, both with high $k$: fundamental
modes with large $n$ and a general resonance with a small $n$. In
addition to finding eigenwavenumbers and showing field plots we also
do a convergence study, compute Q-values, and present far-field
patterns. The convergence study comprises a comparison between our
formulation \eqref{eq:eigenQF} and a homogeneous modal version of the
Müller formulation \eqref{eq:mullercomb2}.

\begin{figure}[!t]
\centering 
\includegraphics[height=51mm]{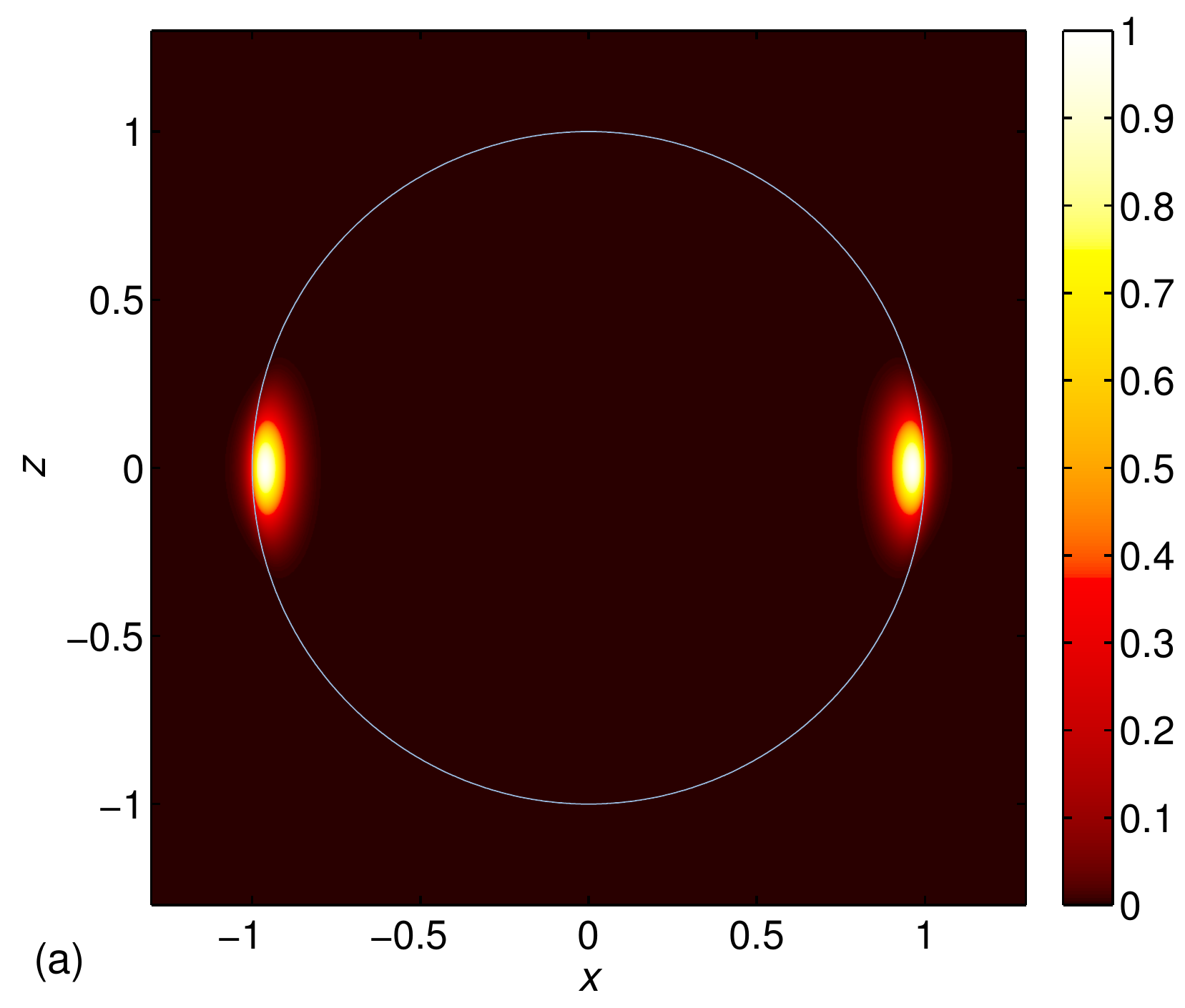}
\includegraphics[height=51mm]{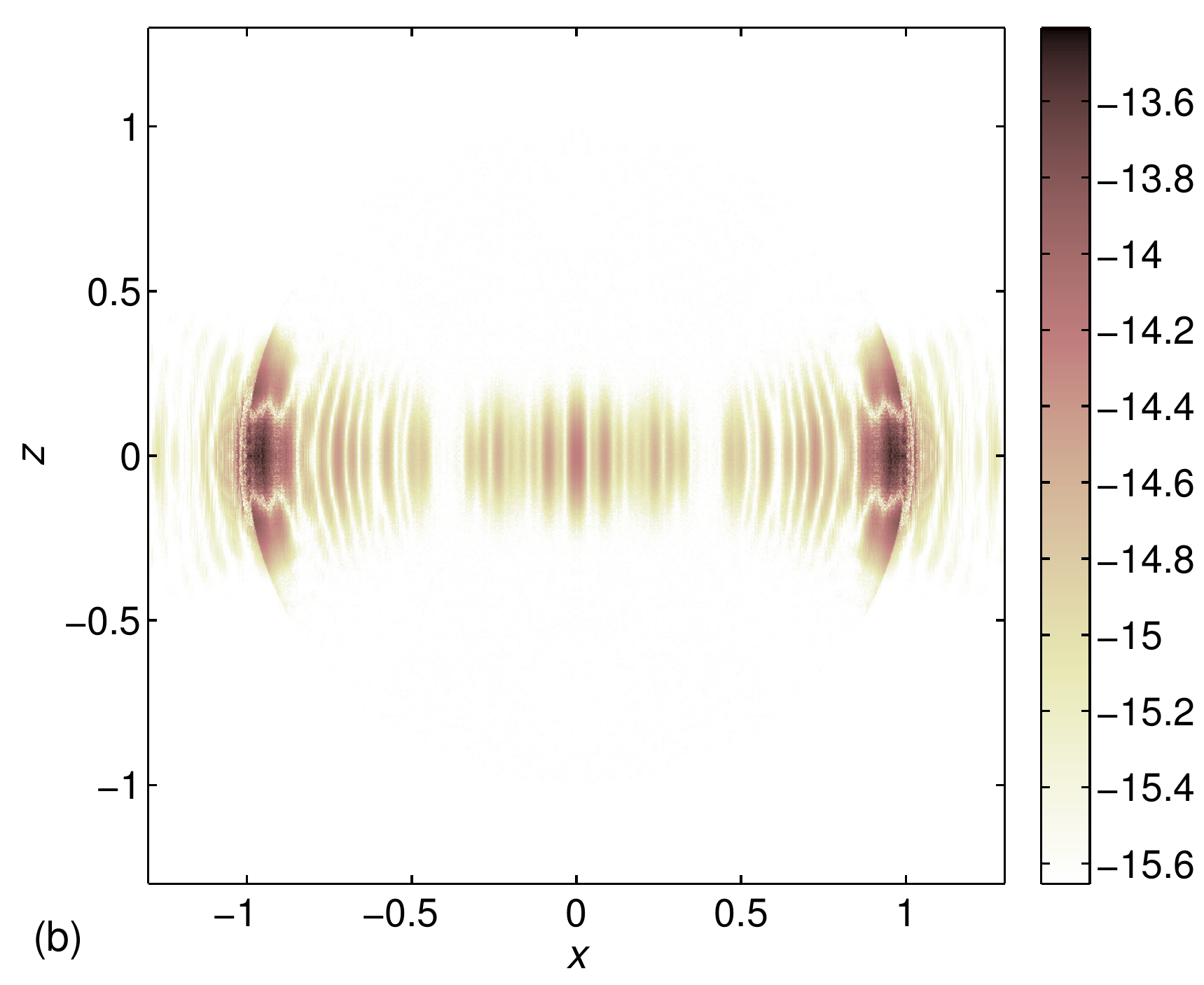}
\includegraphics[height=51mm]{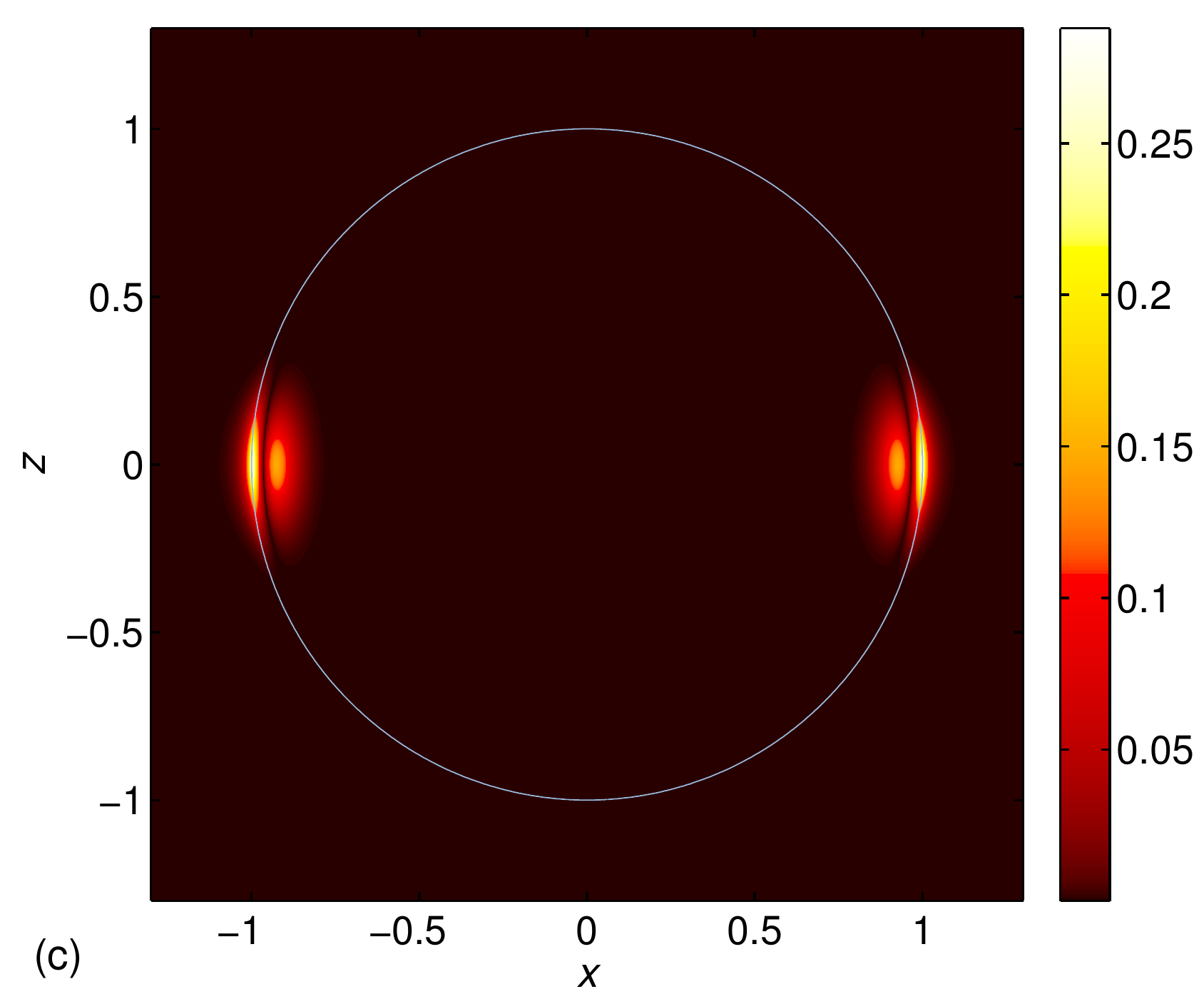}
\includegraphics[height=51mm]{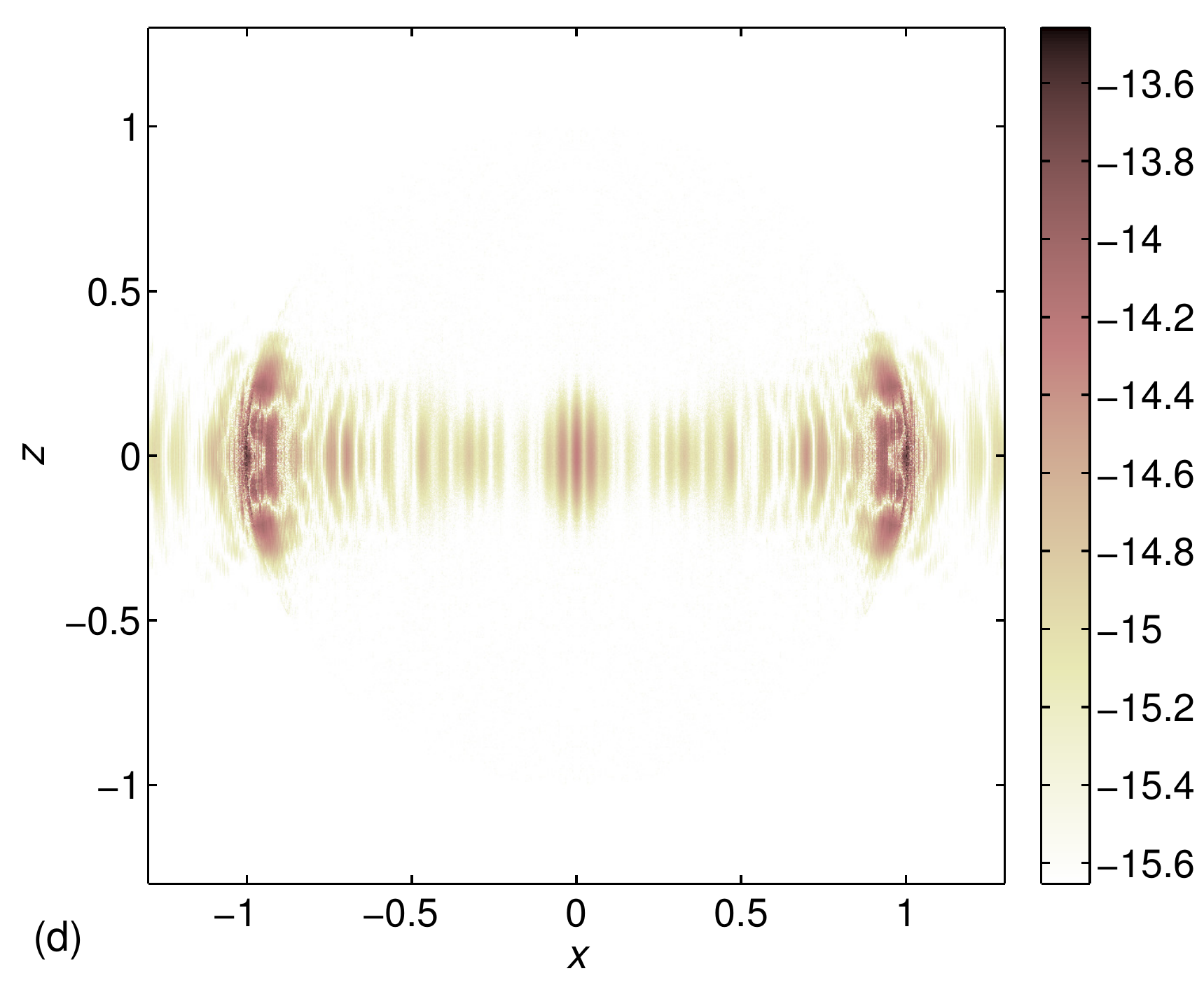}
\includegraphics[height=51mm]{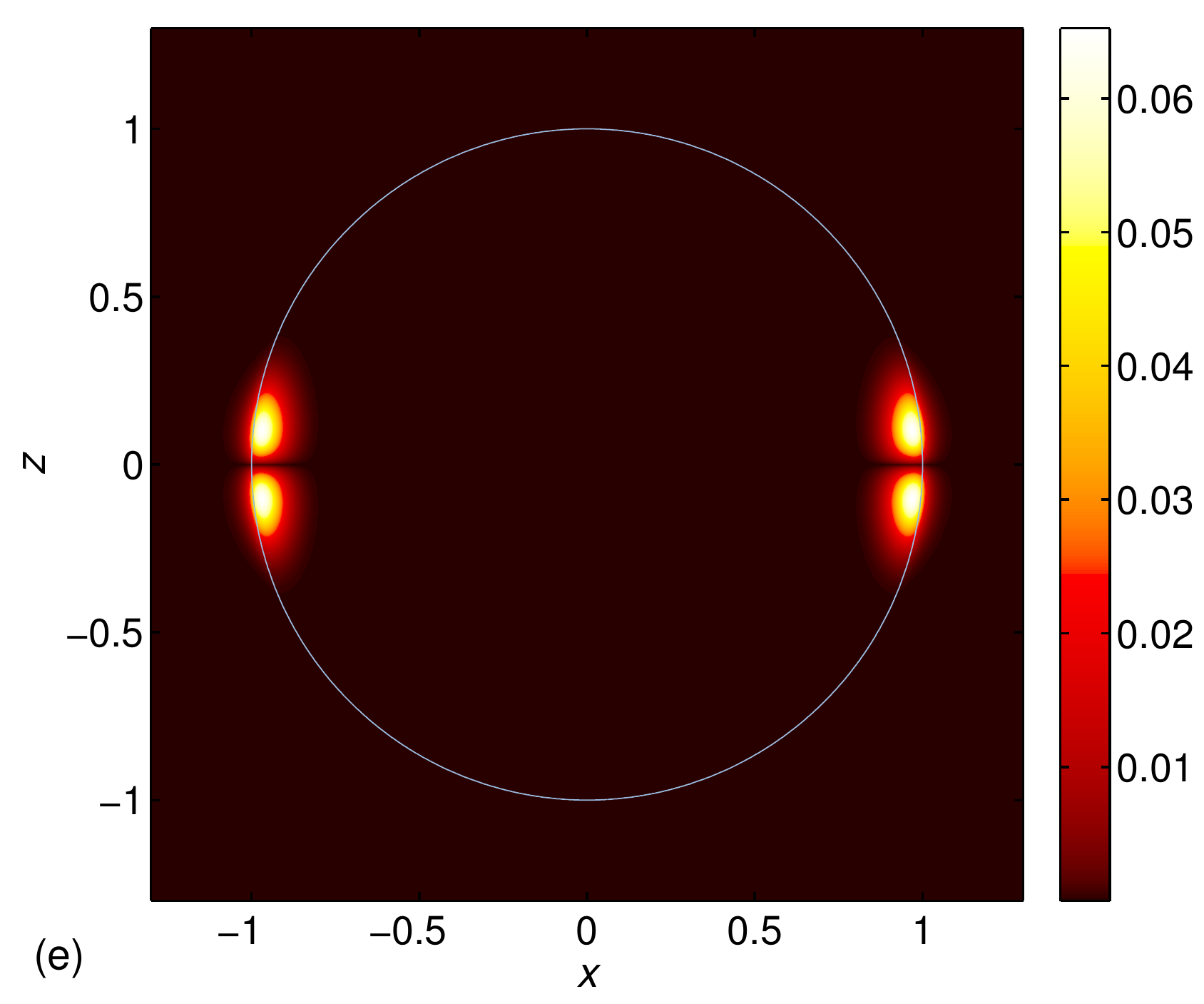}
\includegraphics[height=51mm]{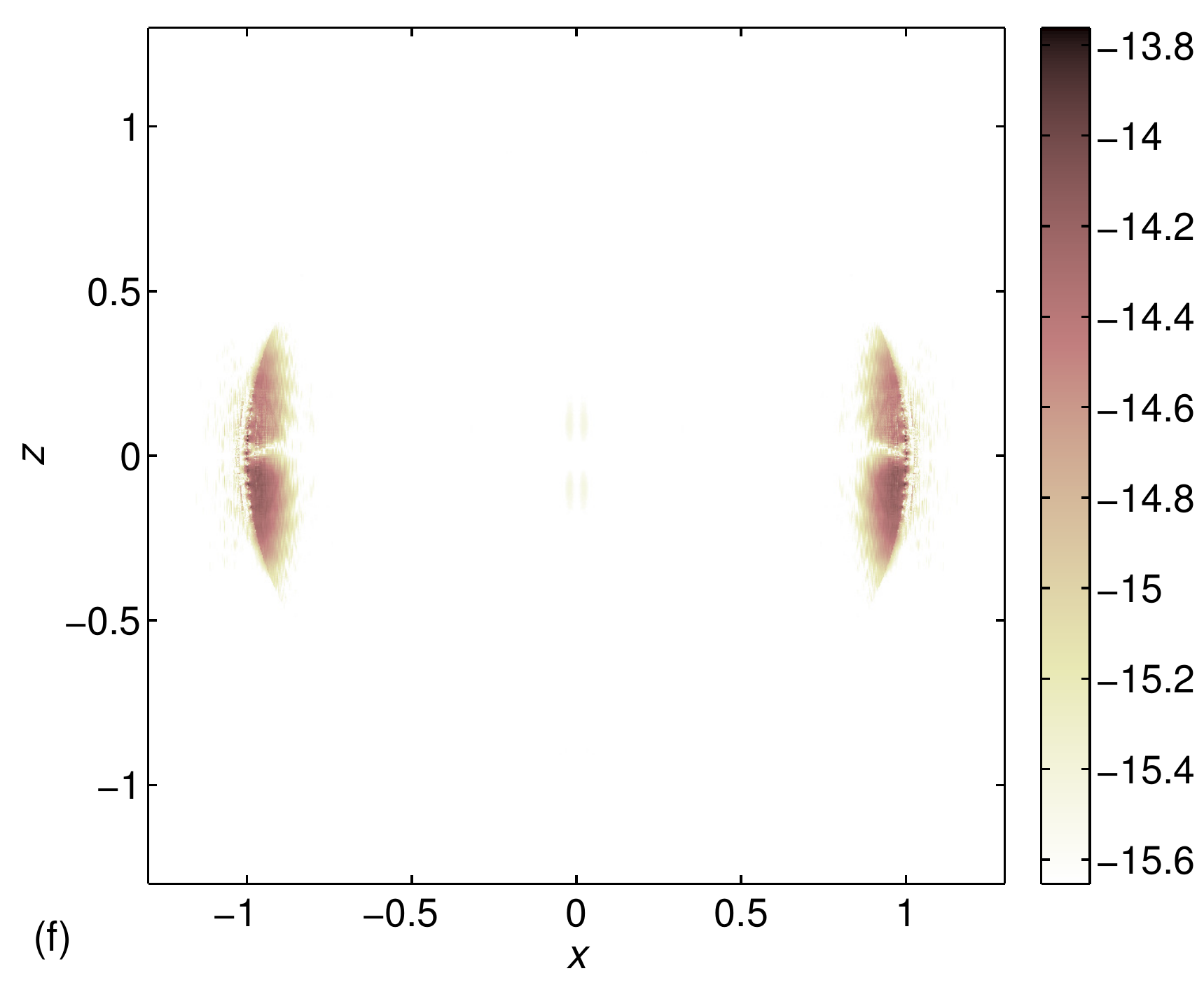}
\caption{\sf Planar field plots of the magnetic field for the
  fundamental $n=90$ mode of a unit sphere with refractive index
  $m=1.5$. The eigenwavenumber is $k=65.09451518155630-1.3\cdot
  10^{-13}{\rm i}$ and $832$ discretization points are used on
  $\gamma$: (a), (c), and (e) show $\vert H_{\rho 90}(r)\vert$, $\vert
  H_{\theta 90}(r)\vert $, and $\vert H_{z90}(r)\vert $; (b), (d), and
  (f) show $\log_{10}$ of the estimated pointwise absolute error.}
\label{fig:n90sph}
\end{figure}

\begin{figure}[!t]
\centering 
\includegraphics[height=51mm]{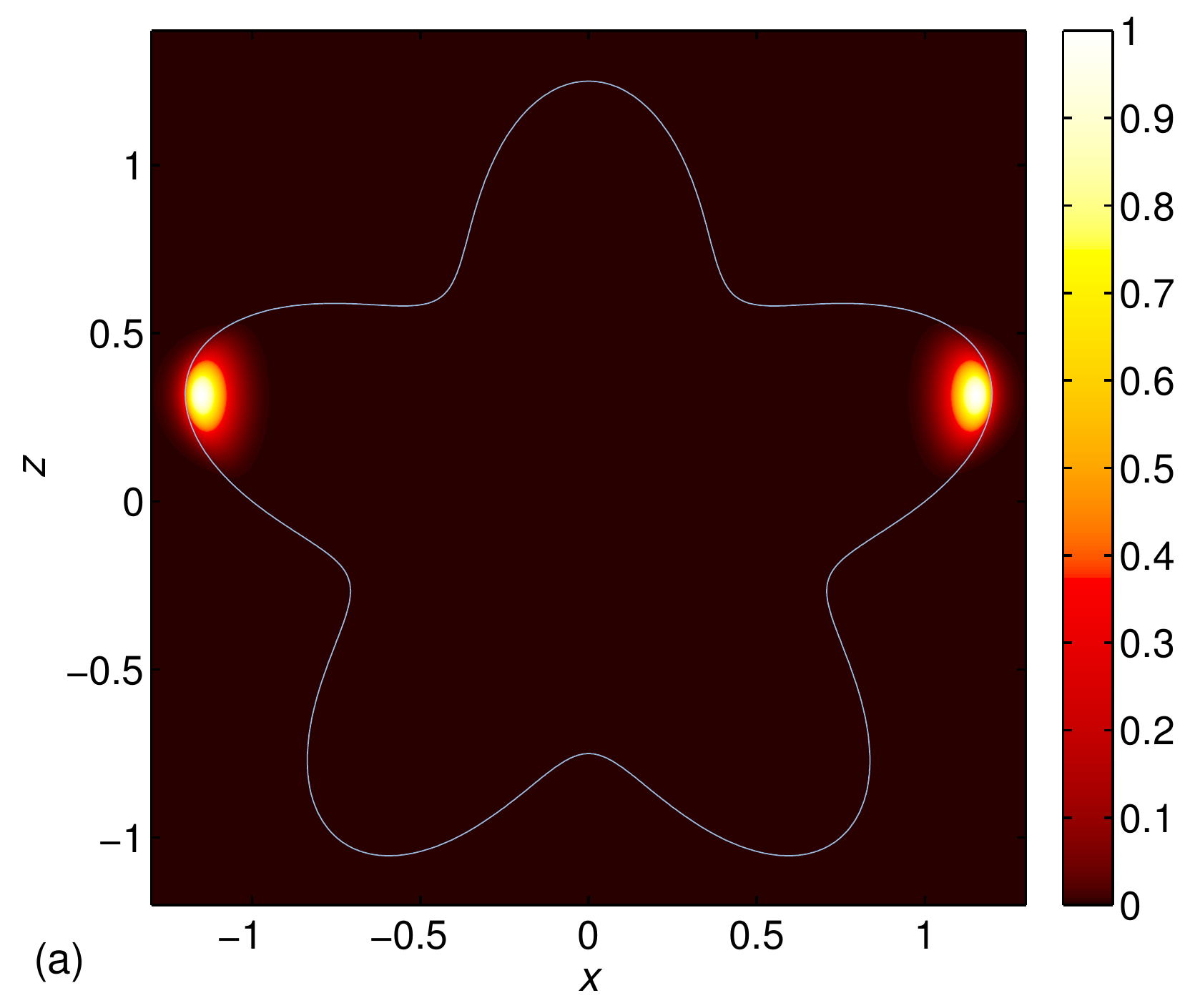}
\includegraphics[height=51mm]{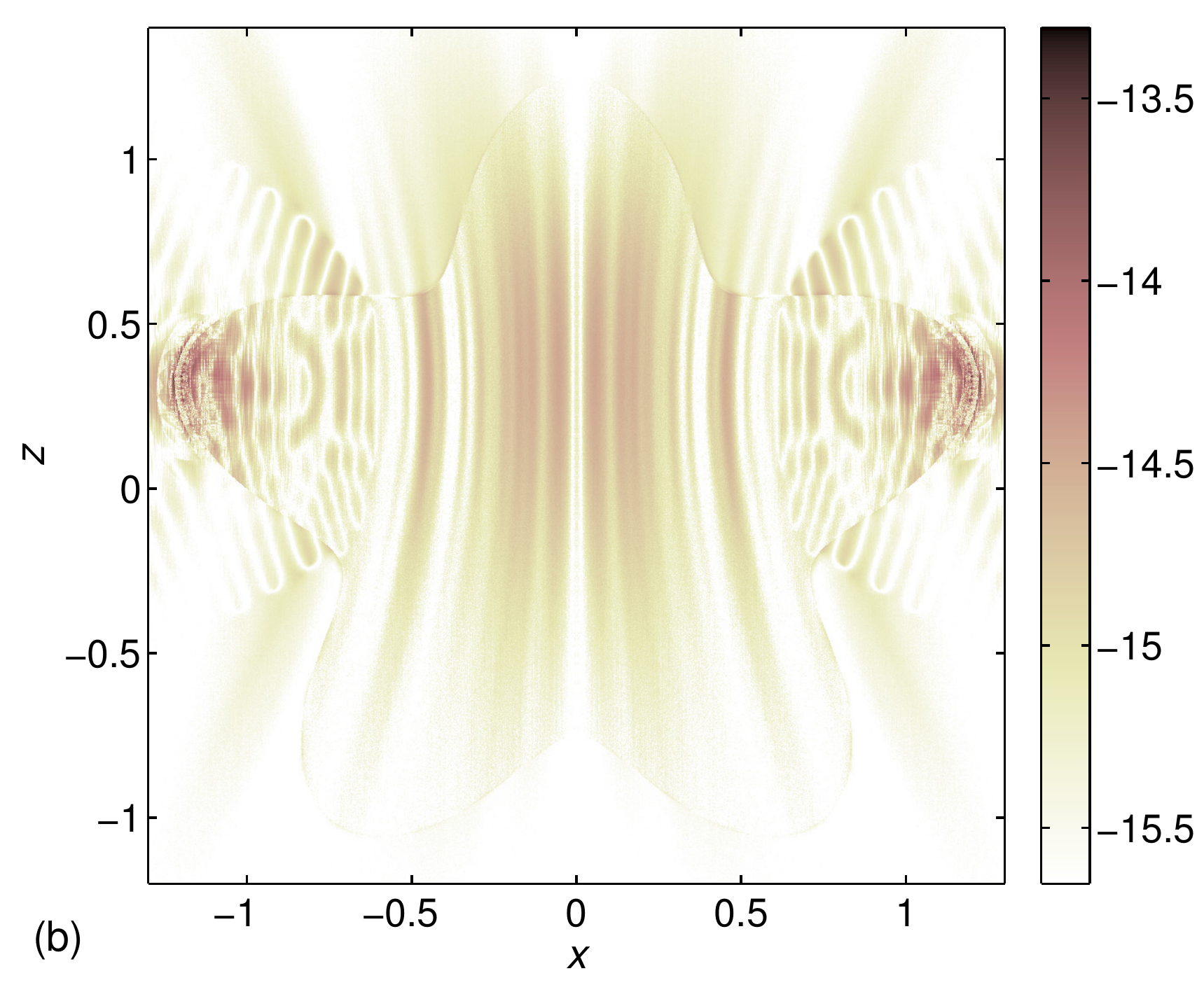}
\includegraphics[height=51mm]{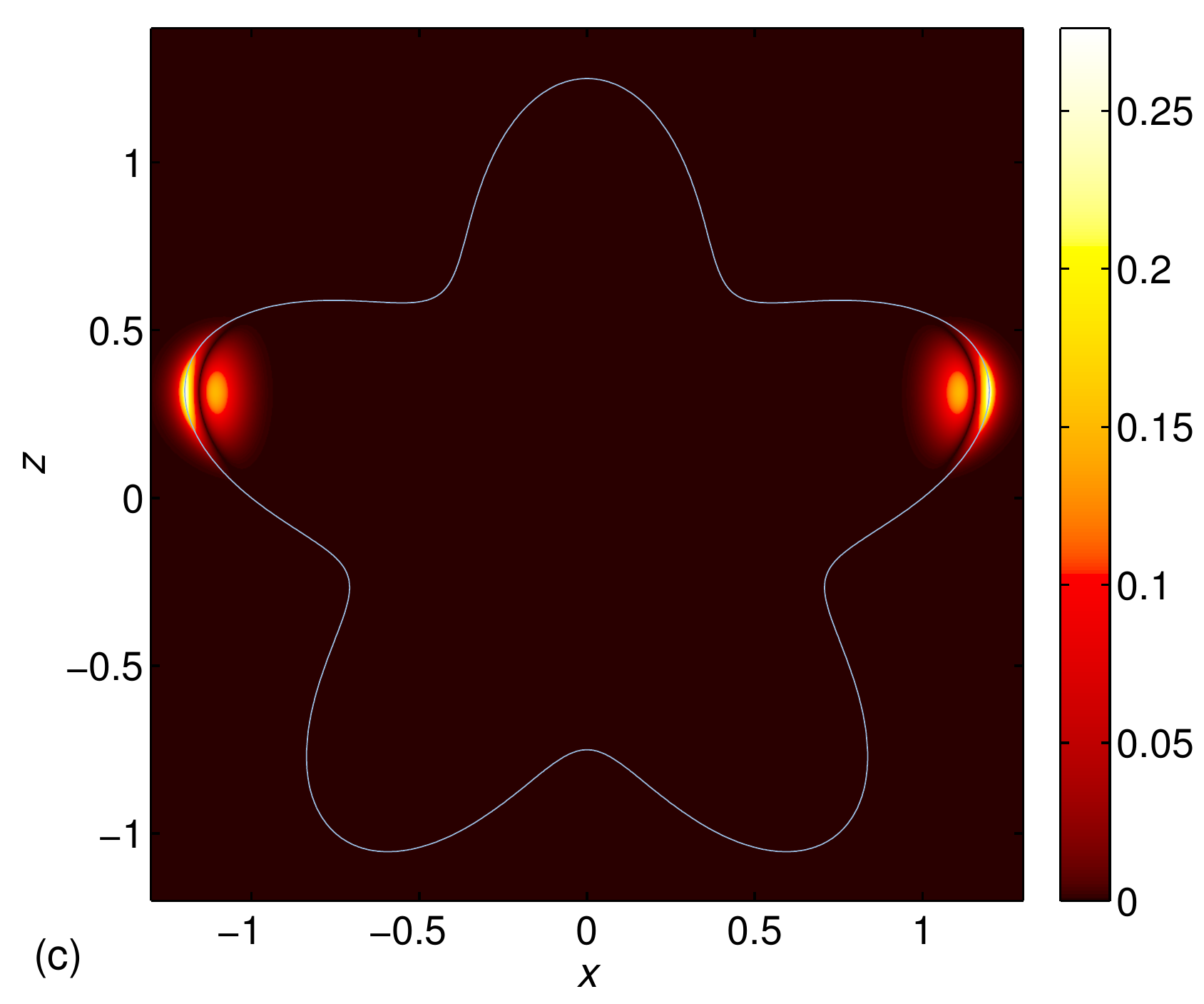}
\includegraphics[height=51mm]{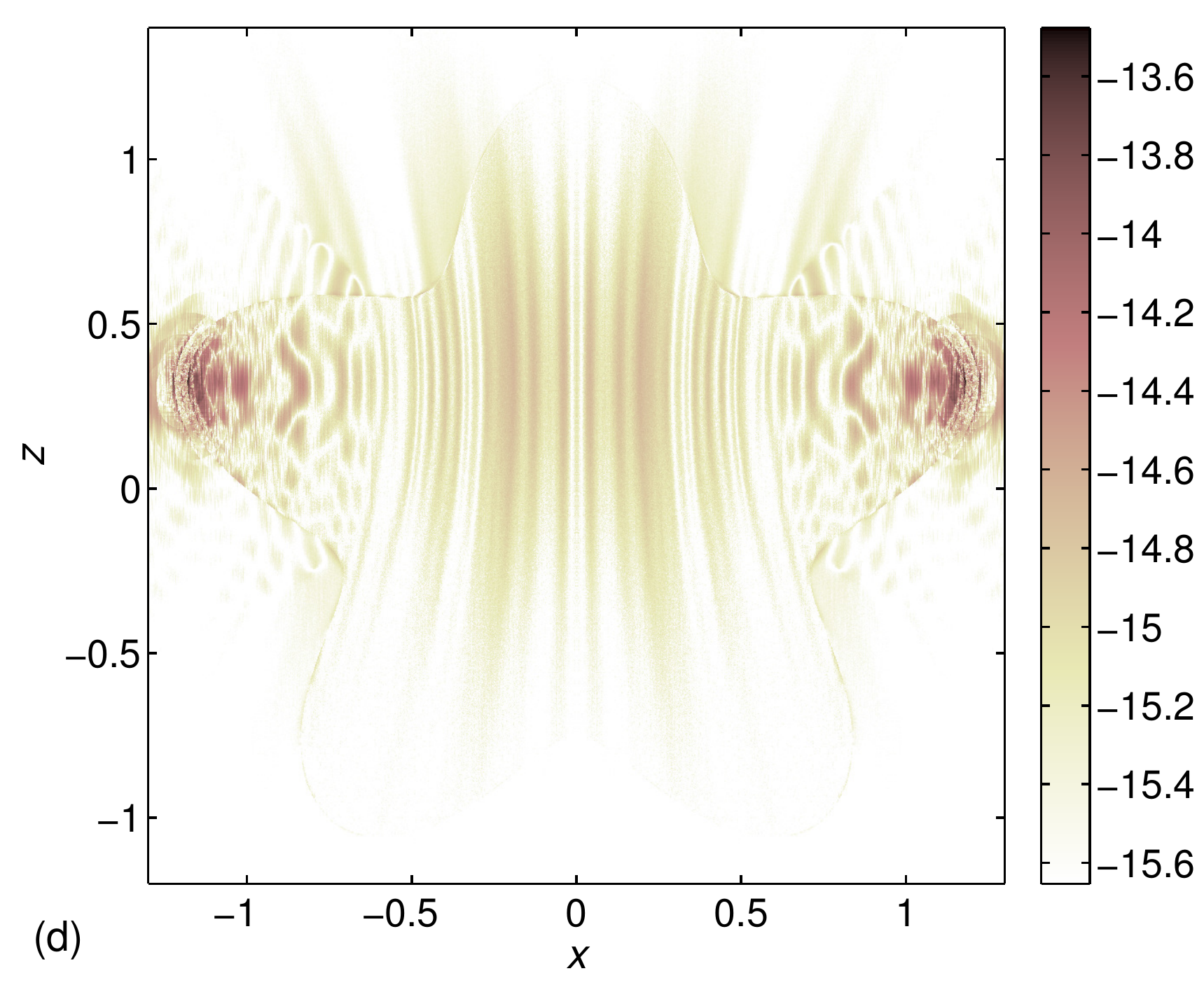}
\includegraphics[height=51mm]{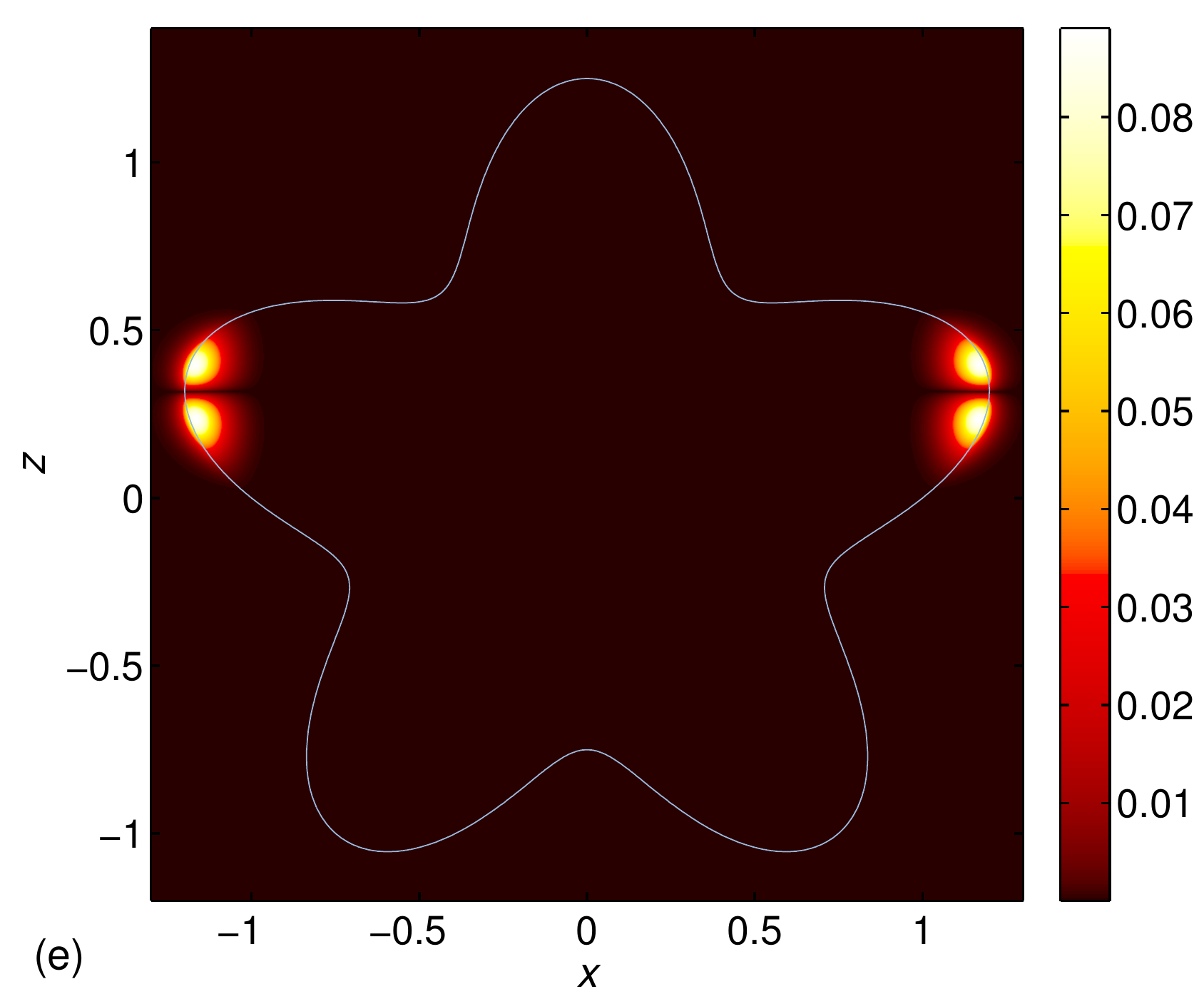}
\includegraphics[height=51mm]{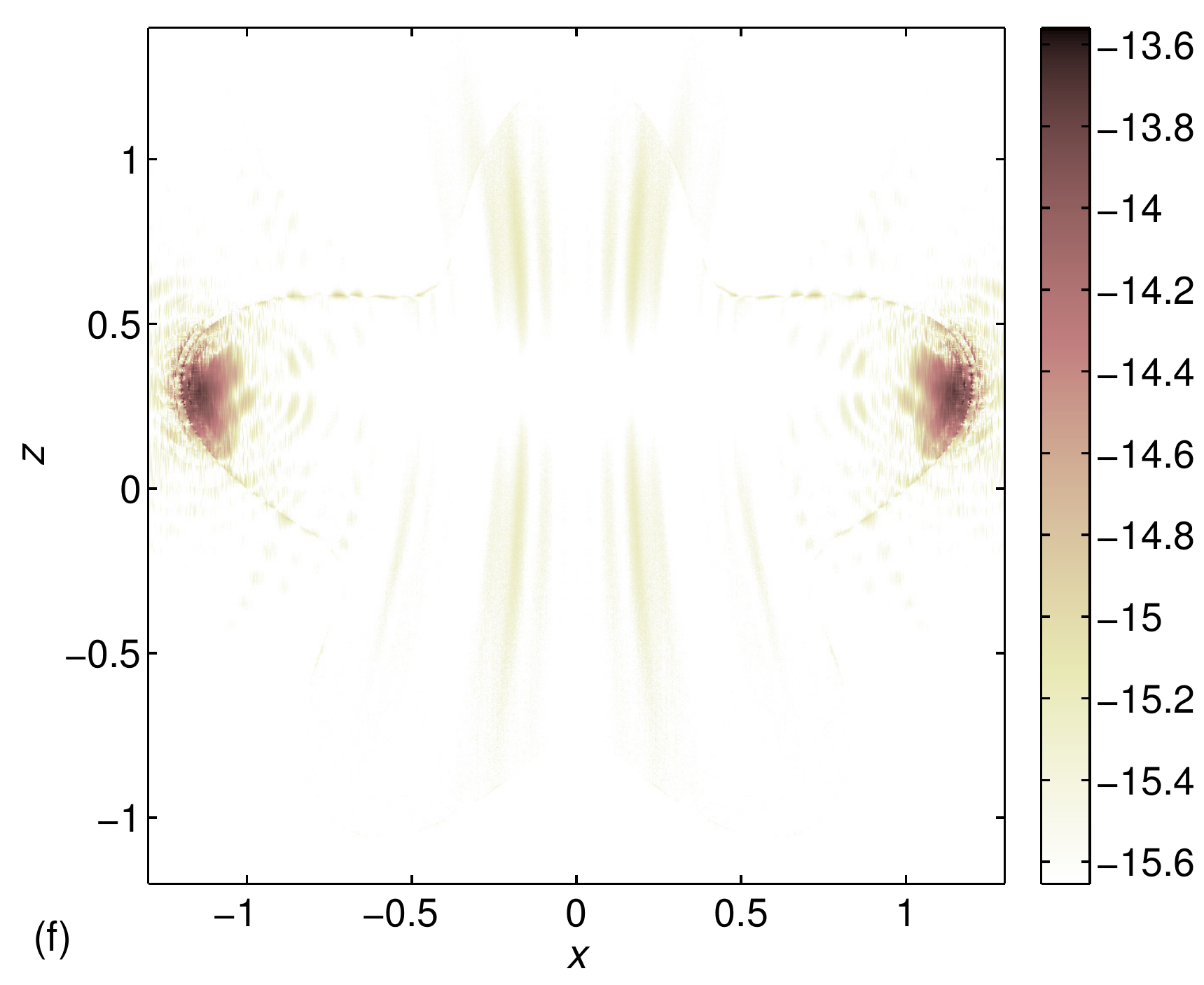}
\caption{\sf Same as in Figure~\ref{fig:n90sph}, but for the object in 
  Figure~\ref{fig:geometry} and with $m=1.5+5.5\cdot 10^{-12}{\rm i}$.
  The eigenwavenumber is $k=54.72590089140112-1.9803\cdot 10^{-10}{\rm
    i}$ and $864$ discretization points are used on $\gamma$.}
\label{fig:n90stfi}
\end{figure}

\begin{figure}[!t]
\centering 
\includegraphics[height=51mm]{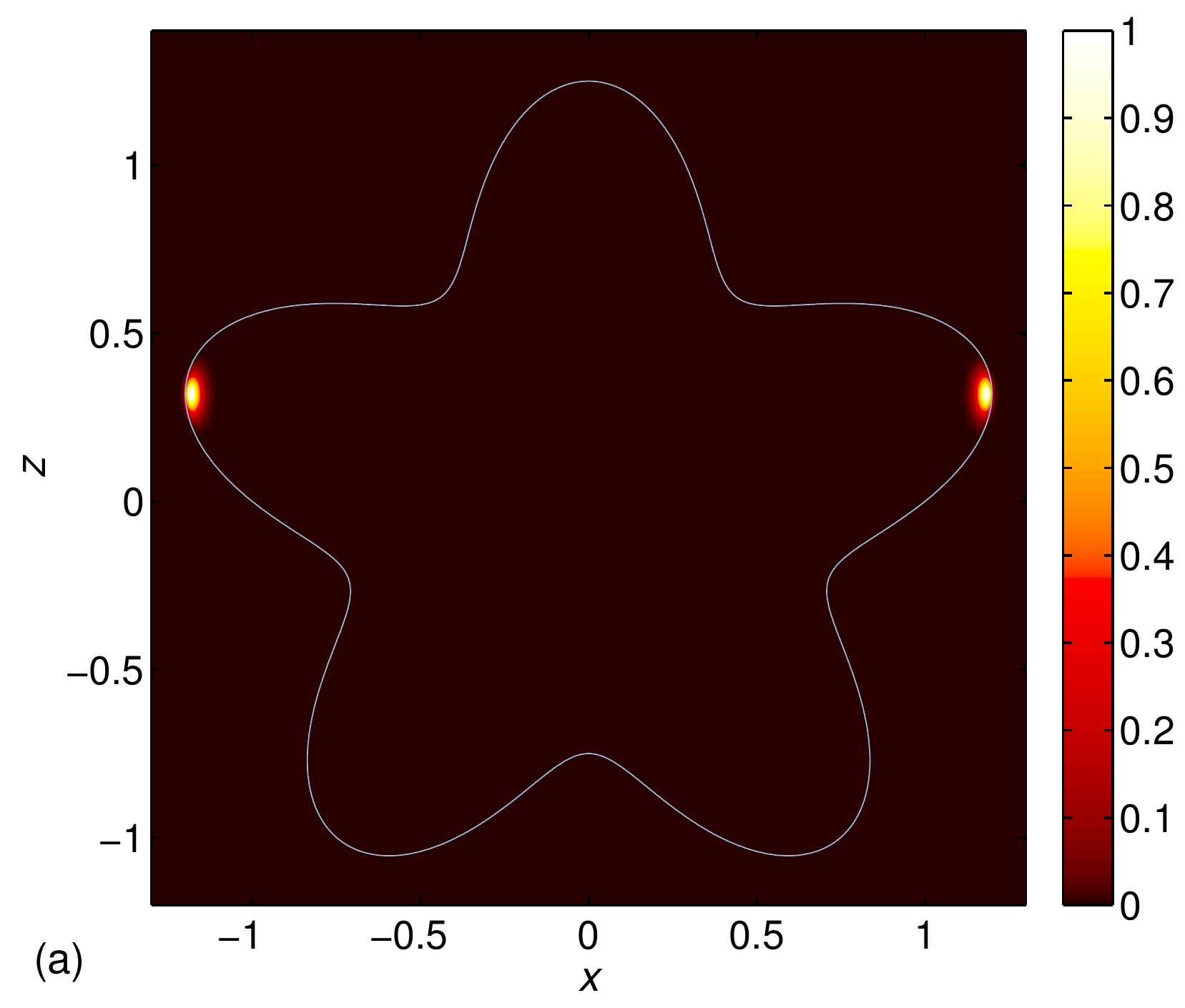}
\includegraphics[height=51mm]{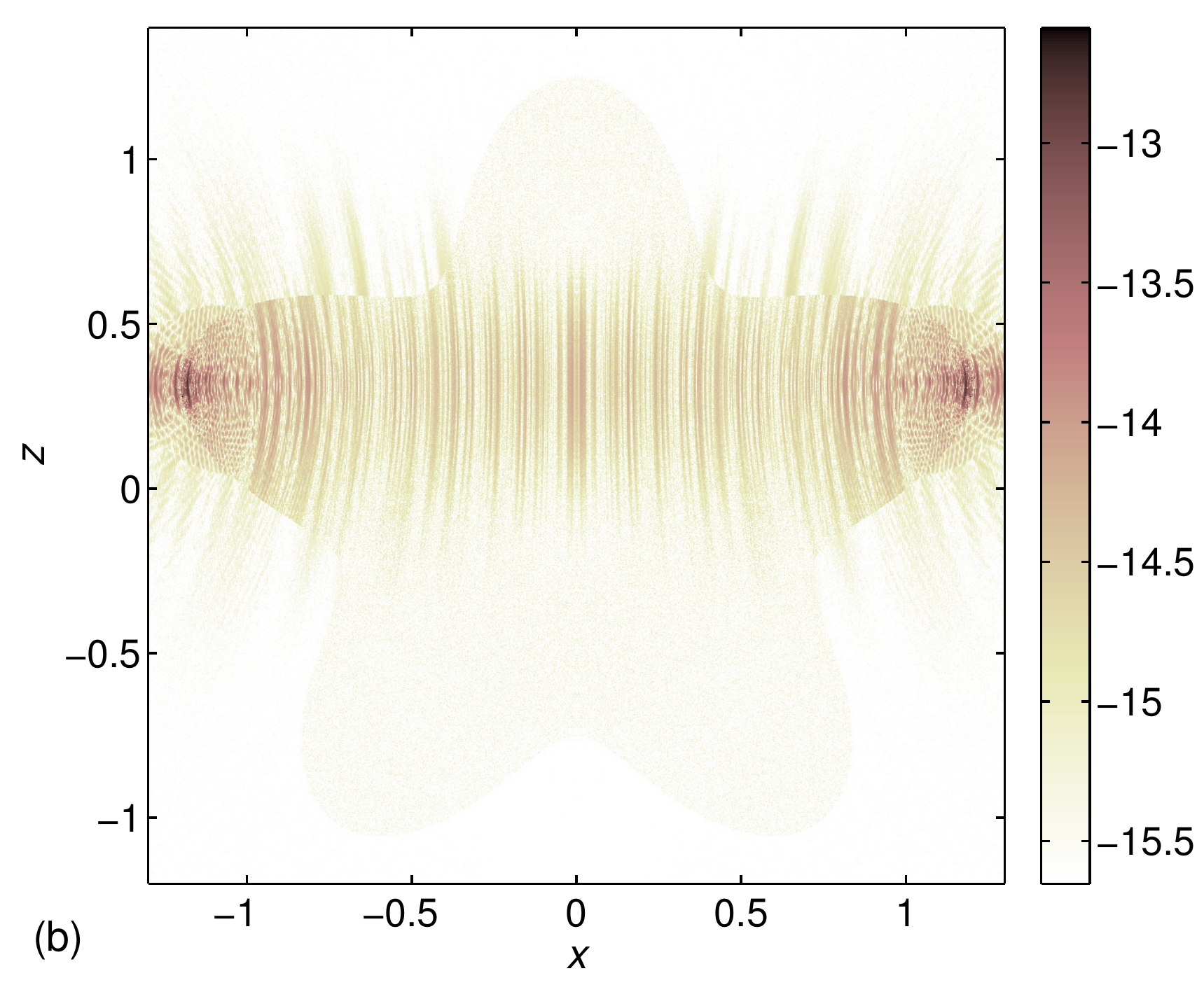}
\includegraphics[height=51mm]{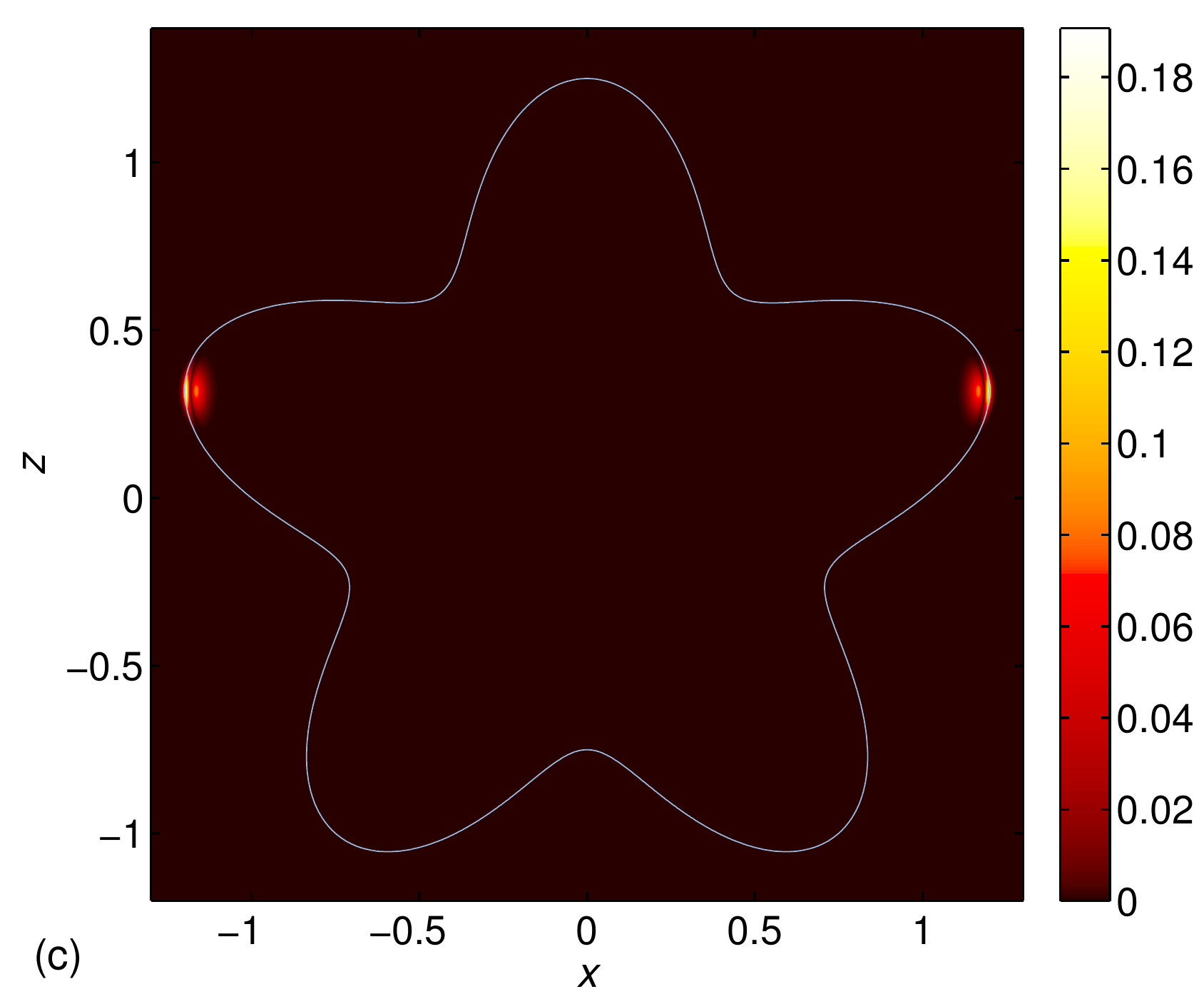}
\includegraphics[height=51mm]{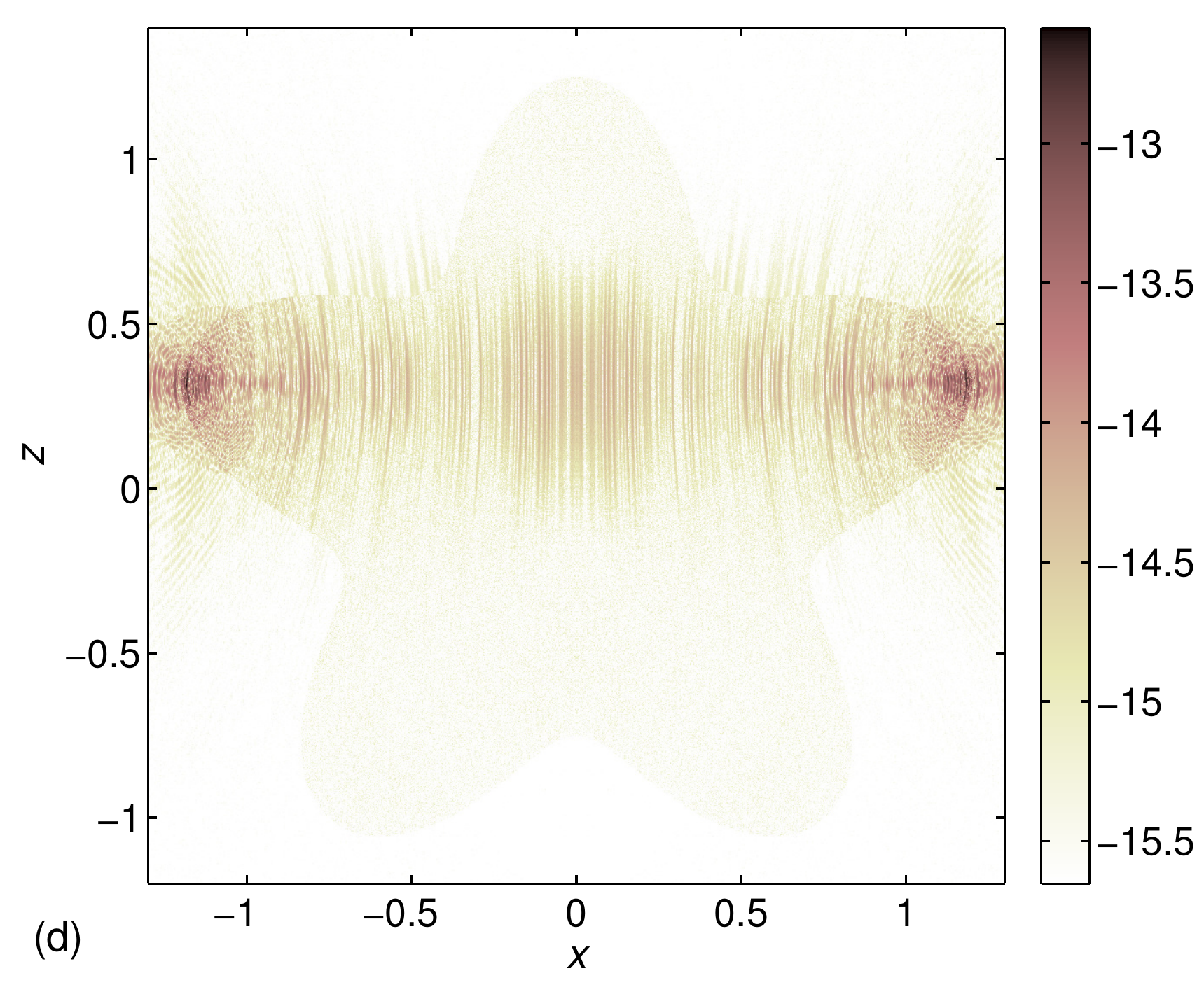}
\includegraphics[height=51mm]{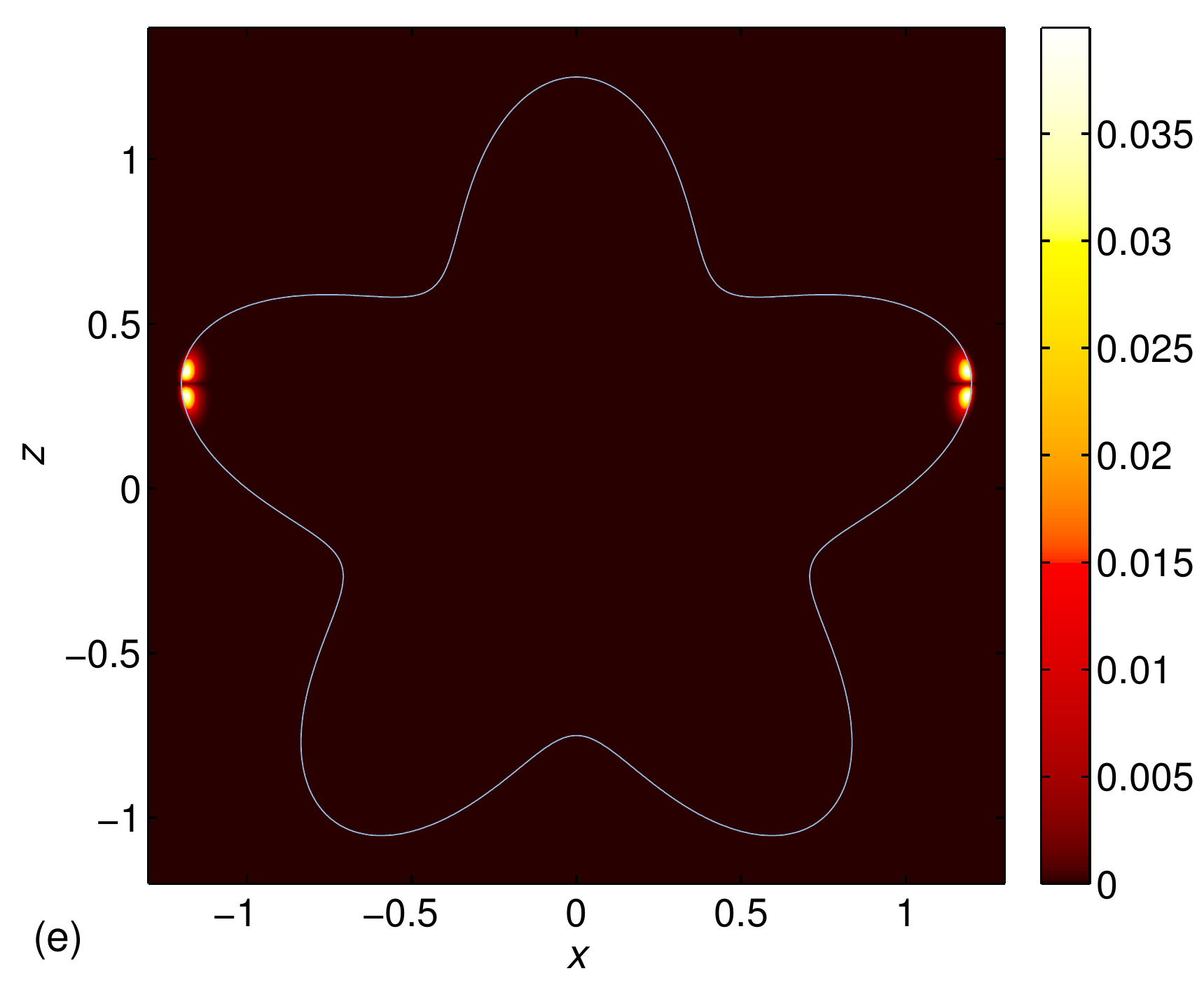}
\includegraphics[height=51mm]{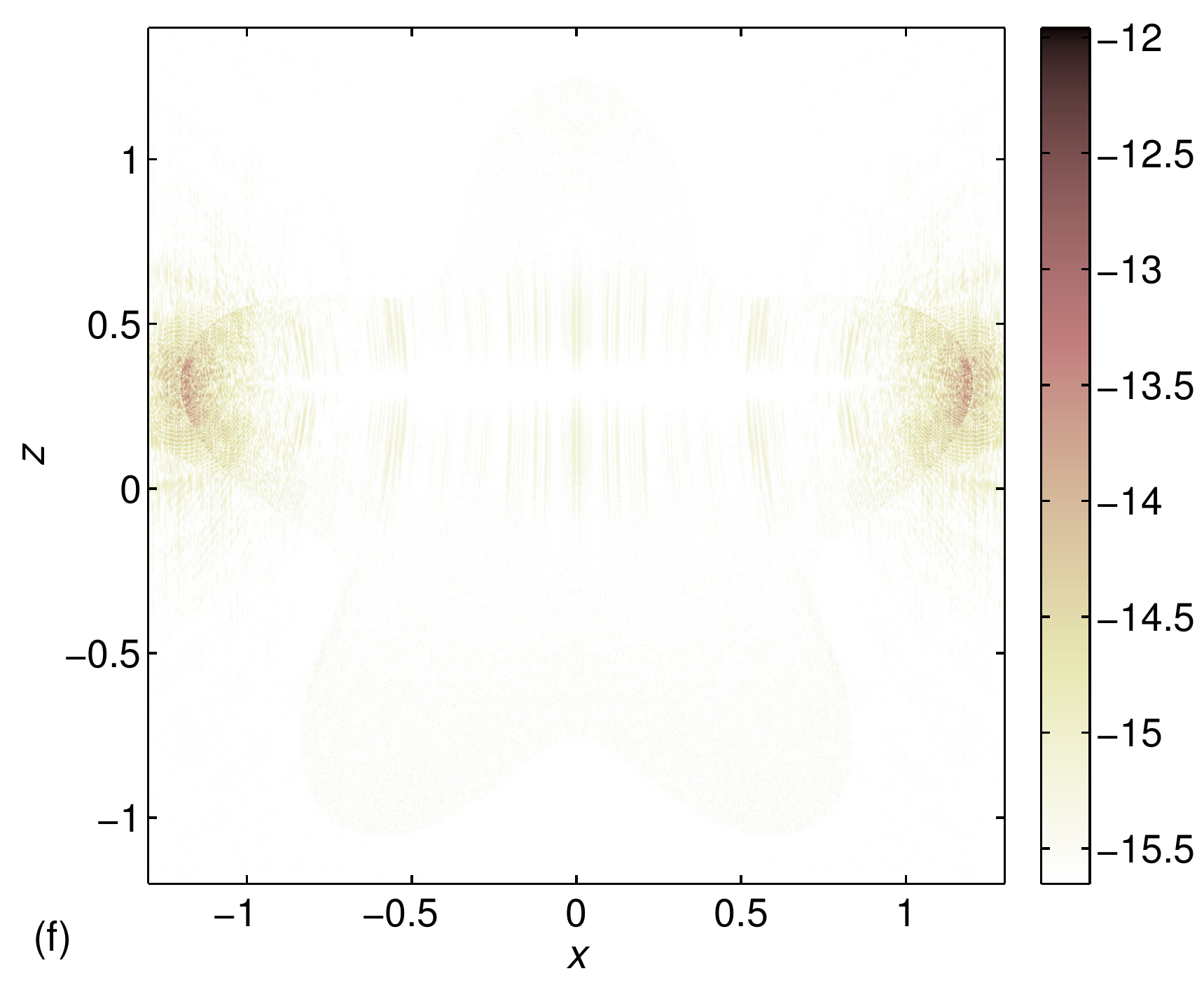}
\caption{\sf Same as in Figure~\ref{fig:n90stfi}, but for $n=450$ and 
  with $m=1.5$. The eigenwavenumber is $k=258.059066513439$ and $352$
  adaptively spaced discretization points are used on $\gamma$.}
\label{fig:n450Hstf}
\end{figure}

\subsection{The fundamental mode for large $n$} 

\subsubsection{The unit sphere}
\label{sec:unit}
  
Our first example is the fundamental $n=90$ mode of the unit sphere
with $m=1.5$ and is intended as a verification of the solver. The
reference solution is evaluated from a semi-analytic solution given by
Mie theory~\cite{Stratton41}. The eigenwavenumber
$k=65.09451518155629-1.3\cdot 10^{-13}{\rm i}$, found by the solver,
corresponds to a sphere diameter of 20.7 vacuum wavelengths and agrees
with the value $k=65.09451518155630-1.3\cdot 10^{-13}{\rm i}$,
obtained from the semi-analytic solution, to almost machine precision.

Figure~\ref{fig:n90sph} shows field plots of $(\vert H_{\rho
  90}(r)\vert ,\vert H_{\theta 90}(r)\vert ,\vert H_{z90}(r)\vert )$
along with estimated absolute pointwise errors, which peak at around
$100\epsilon_{\rm mach}$. In passing we mention that our numerical
tests revealed the following relations for the fundamental modes of
dielectric spheres:
\begin{equation}
E_{\theta n}(r)
={\rm i}E_{\rho n}(r)
={\rm i}\dfrac{z}{\rho}E_{zn}(r)\,,
\end{equation}
which we then also derived from the semi-analytic solution.

\subsubsection{Lossless versus lossy object materials}
\label{sec:LvsL}

We now look at the fundamental $n=90$ mode of the object in
Figure~\ref{fig:geometry} and compare converged eigenwavenumbers and
eigenfields for two different object materials. The first material is
lossless with $m=1.5$. The eigenwavenumber is
$k=54.72590089140112-1.5\cdot 10^{-13}{\rm i}$, corresponding to a
generalized object diameter of about 22.8 vacuum wavelengths. The
Q-factor \eqref{eq:Q} is ${\rm Q}=1.8\cdot 10^{14}$. The second
material is lossy with $m=1.5+5.5\cdot 10^{-12}{\rm i}$ and has
$k=54.72590089140112-1.9803\cdot 10^{-10}{\rm i}$, which corresponds
to a skin depth $\delta\approx 10^{10}$. The
condition~(\ref{eq:skindepth}) is fulfilled and by that ${\rm Q}_{\rm
  rad}$ is independent of $\Im{\rm m}\{m\}$ and \eqref{eq:Qapp} holds.
A comparison of the eigenwavenumbers reveals that $\Re{\rm e}\{k\}$ is
virtually unaffected by the losses. Since ${\rm Q}_{\rm rad}\gg {\rm
  Q}_{\rm diss}$, it follows from~(\ref{eq:Qfactors}), (\ref{eq:Q}),
and (\ref{eq:Qapp}) that $\Im{\rm m}\{k\}/\Re{\rm
  e}\{k\}\approx-\Im{\rm m}\{m\}/\Re{\rm e}\{m\}$. Now ${\rm
  Q}=1.382\cdot 10^{11}$ and, since the lossless material has ${\rm
  Q}=1.8\cdot 10^{14}$, the dissipative Q-factor is ${\rm Q}_{\rm
  diss}=1.383\cdot 10^{11}$ according to \eqref{eq:Qfactors}. This
value agrees well with the approximate expression ${\rm Q}_{\rm
  diss}\approx 1.364\cdot 10^{11}$ from~(\ref{eq:Qapp}).

The losses of the second material is the same as that of silica at the
vacuum wavelength of 1550 nm, which is the smallest known loss of any
solid material at optical wavelengths. It indicates that the physical
limit for the Q-factor is approximately $10^{11}$. To the eye, the
field plots and the corresponding error plots with the lossless
material and with the lossy material are indistinguishable. The images
shown in Figure~\ref{fig:n90stfi} are thus valid for both object
materials. We also tested our scheme for the large loss case with
$m=1.5+5\cdot 10^{-3}{\rm i}$, giving
$k=54.72532533461791-0.17989547170087{\rm i}$, without any problem.

\subsubsection{A high wavenumber WGM}
\label{sec:hiWGM}

Figure~\ref{fig:n450Hstf} shows planar plots and error estimates of
the magnetic field for the fundamental $n=450$ mode of the object in
Figure~\ref{fig:geometry} with $m=1.5$. The eigenwavenumber of this
WGM is $k=258.059066513439$, corresponding to a generalized object
diameter of about 107.6 vacuum wavelengths. This is in the regime
where asymptotic methods for WGMs are applicable~\cite{Breu13}. The
imaginary part of $k$ is not identically zero, but it is too small to
be resolved in double precision arithmetic. The images in
Figure~\ref{fig:n450Hstf} resemble those of the fundamental $n=90$
mode in Figure~\ref{fig:n90stfi}, but with the fields confined to a
smaller region and one digit of precision lost. This case has been
compared with an evaluation in COMSOL Multiphysics 5.2, which is a FEM
simulation package.  Since the field is confined to a small region it
was possible to reduce the computational domain in COMSOL to a square
0.25$\times$0.25. By that the eigenwavenumber and the eigenfield could
be evaluated using default meshes. The convergence of the wavenumber
evaluation was of order 1.41. The default mesh referred to as
extremely fine used 2557 degrees of freedom and gave the
eigenwavenumber 258.1164 with a relative error of 2.2$\cdot10^{-4}$.
All of our attempts to analyse other resonances than WGMs with FEM
methods failed. This is in contrast to our experience from perfectly
conducting cavities, \cite{HelsKarl15} and \cite{HelsKarl16}, where
FEM is an option also for other resonances.

\begin{figure}[!t]
\centering 
\includegraphics[height=51mm]{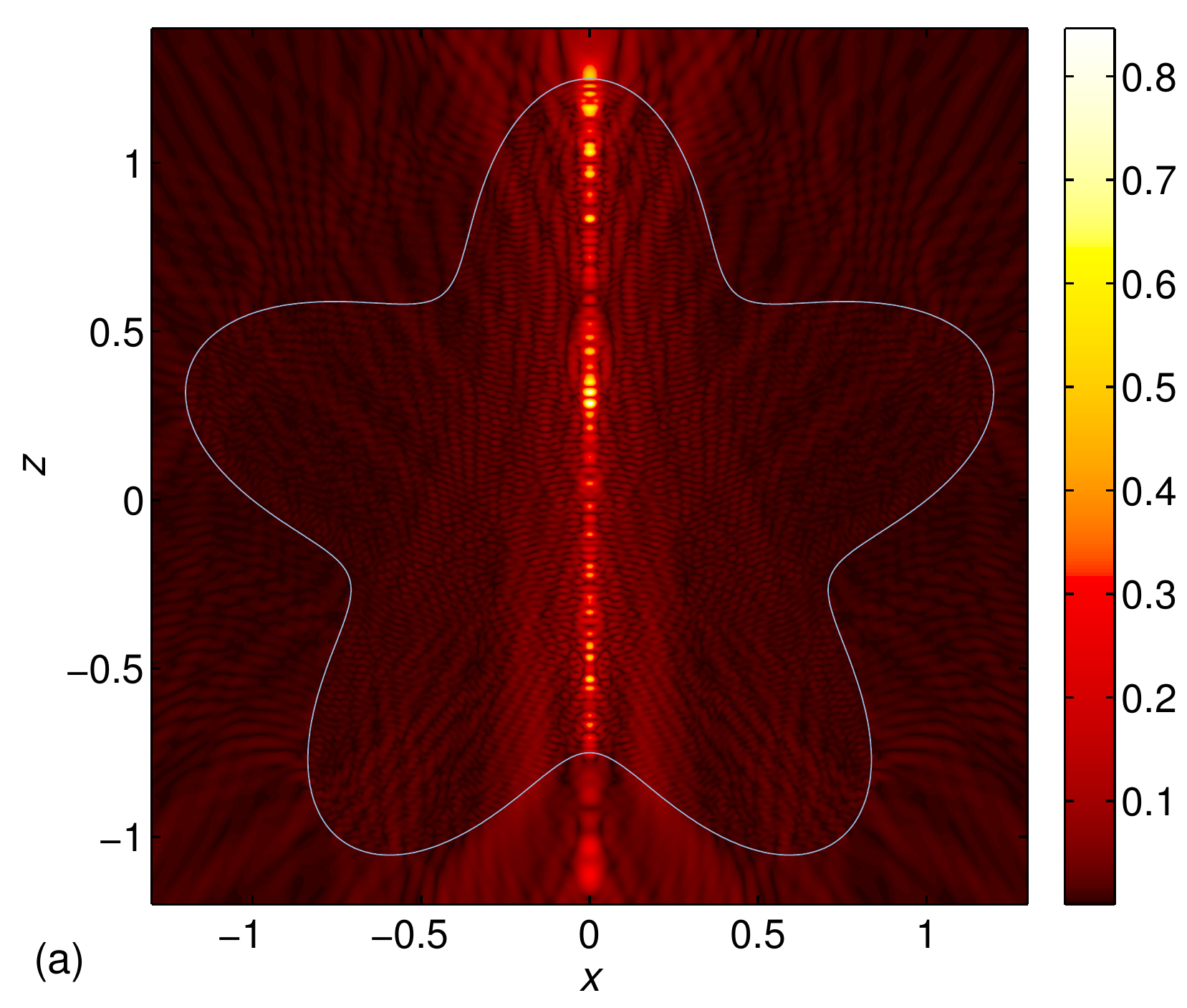}
\includegraphics[height=51mm]{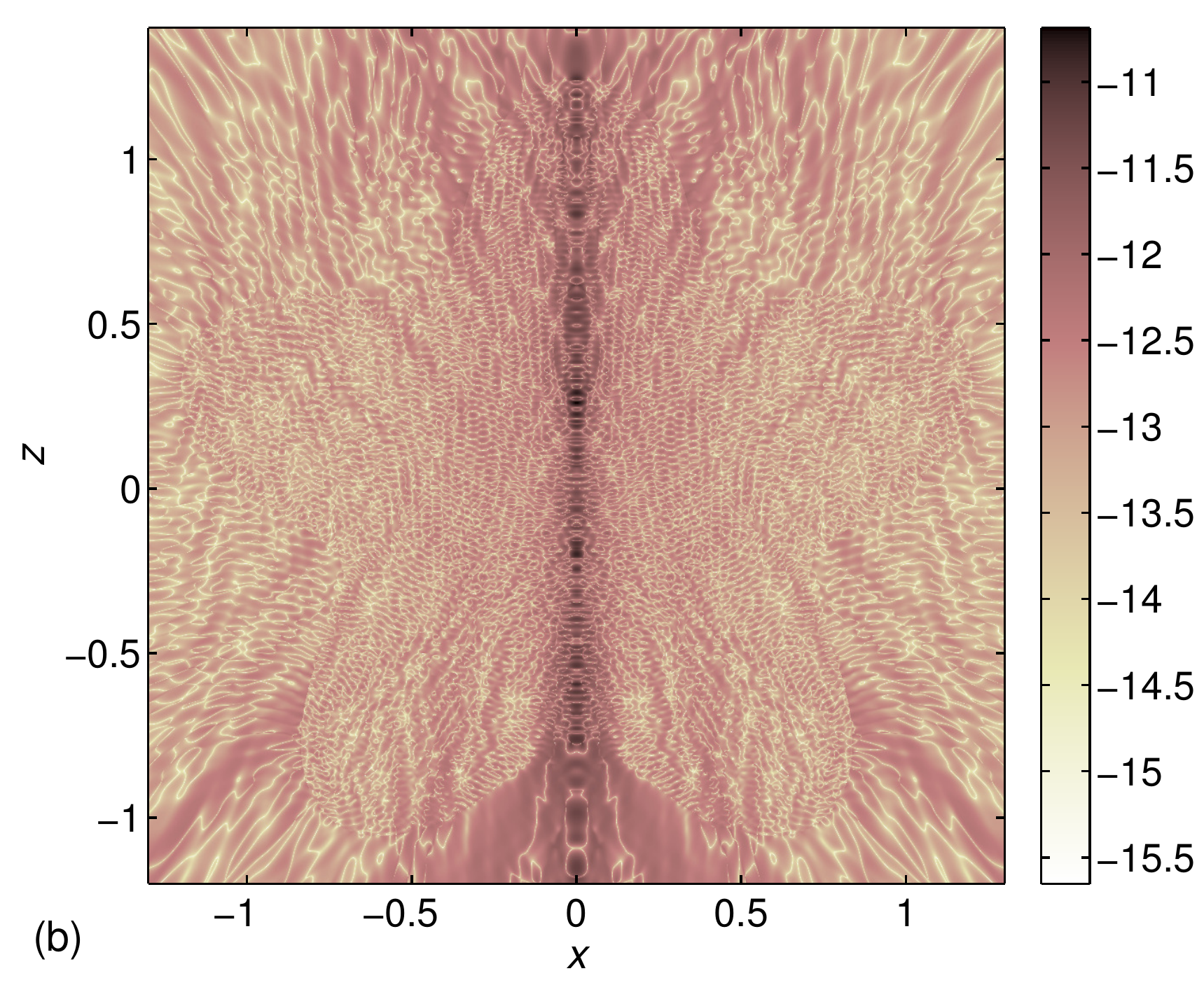}
\includegraphics[height=51mm]{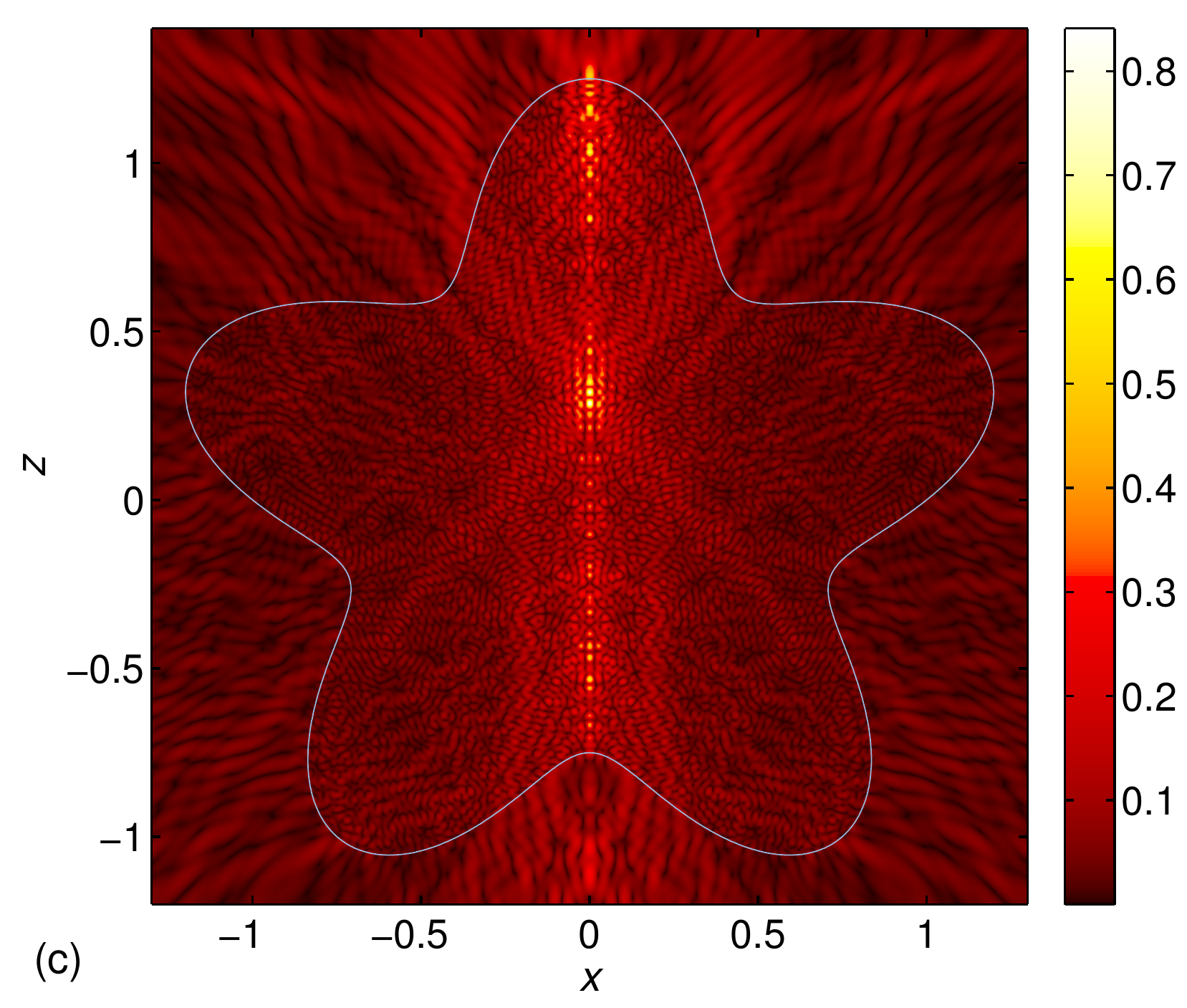}
\includegraphics[height=51mm]{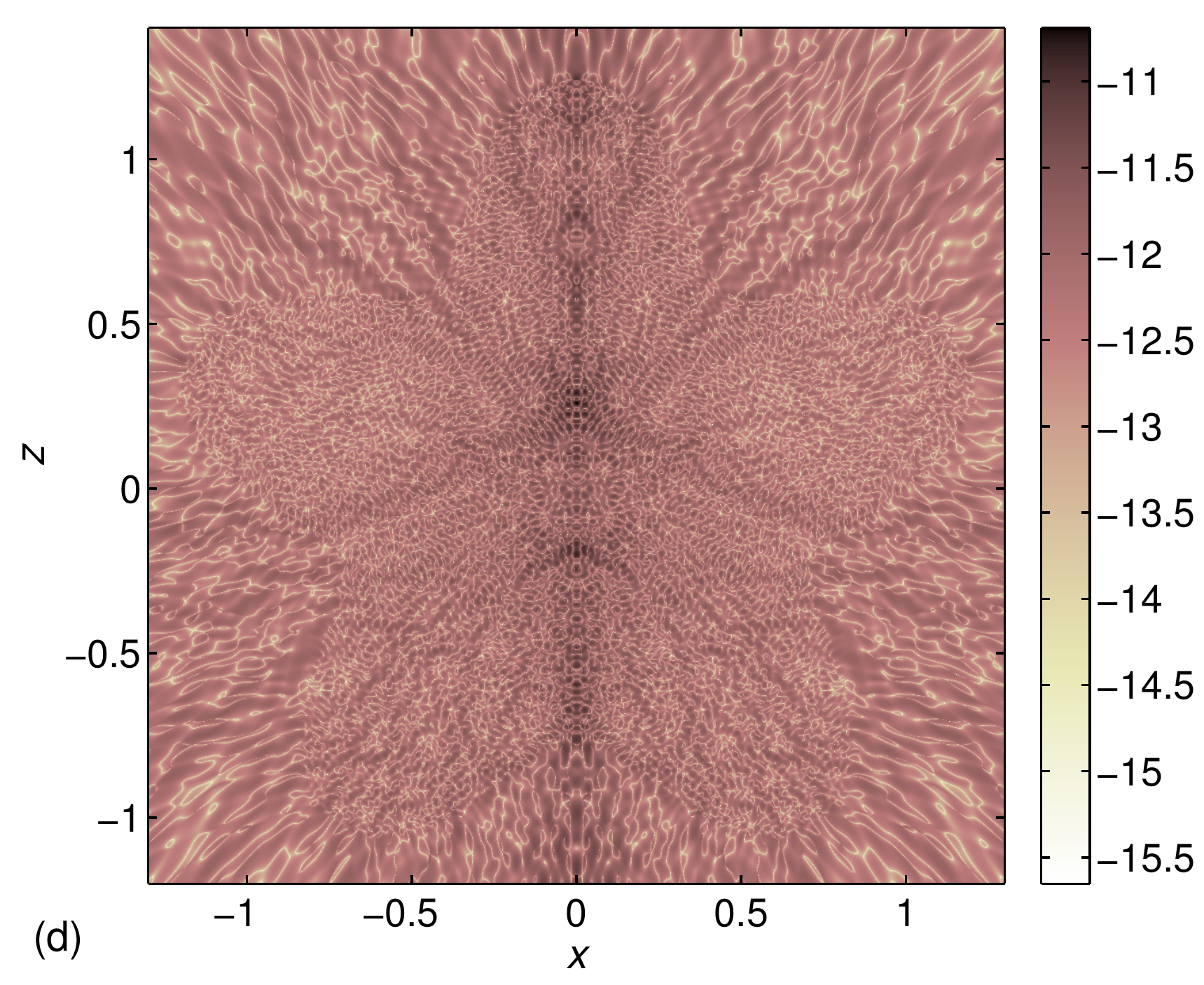}
\includegraphics[height=51mm]{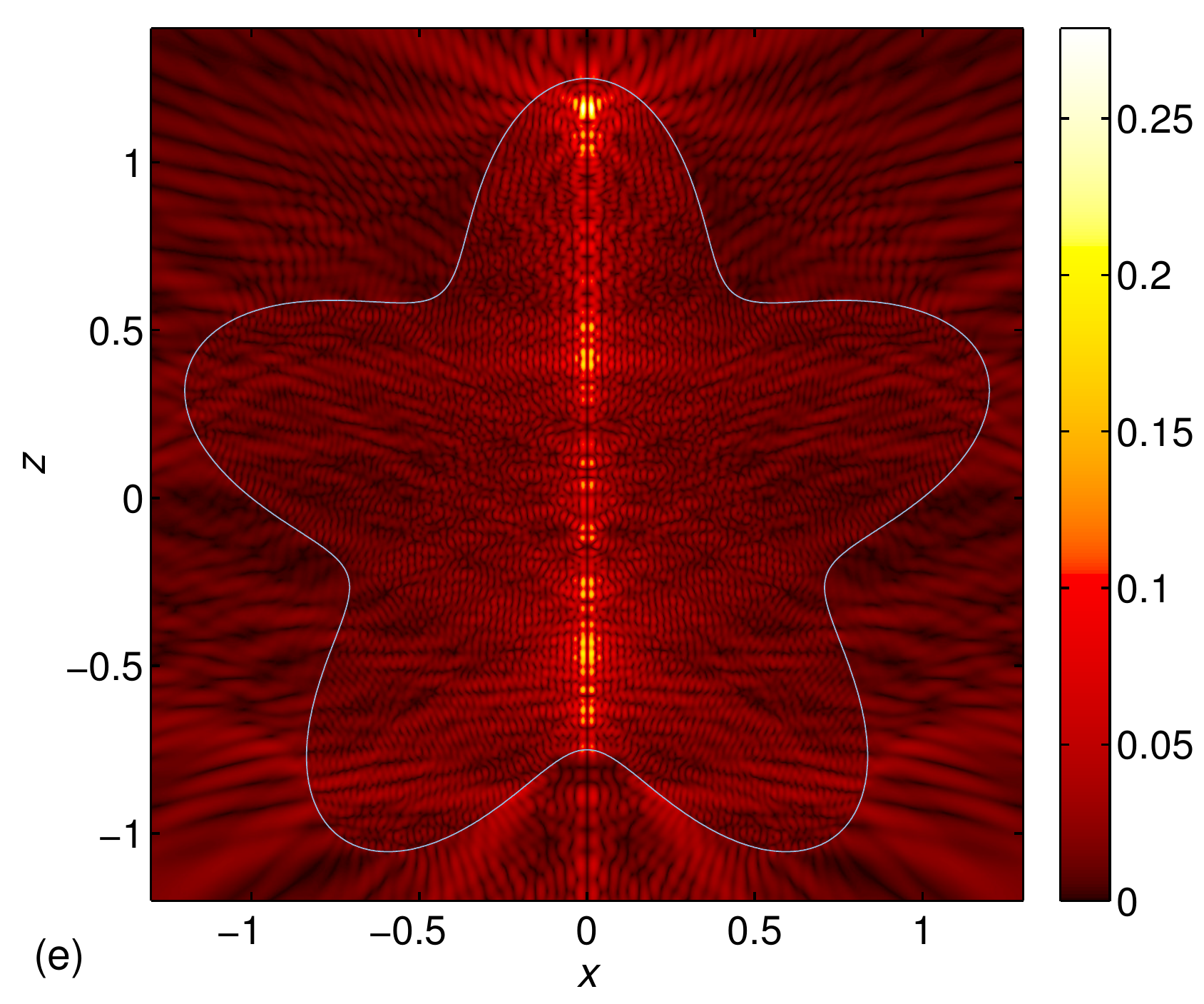}
\includegraphics[height=51mm]{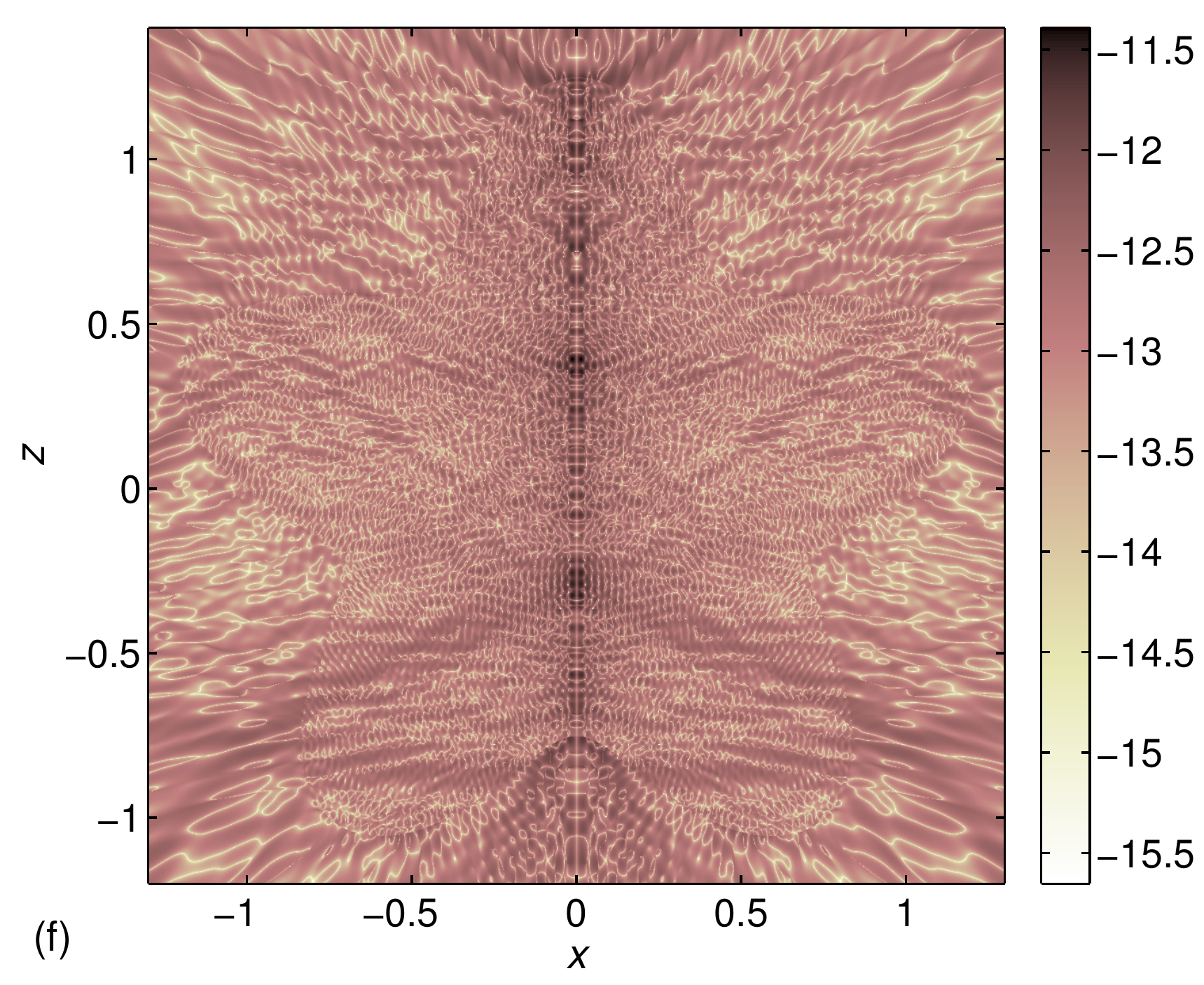}
\caption{\sf Planar field plots of the electric field for an $n=1$ 
  mode of the object in Figure~\ref{fig:geometry}. The eigenwavenumber
  is $k=110.041232211051-0.404177078290{\rm i}$, the refractive index
  is $m=1.5$, and $1984$ discretization points are used on $\gamma$:
  (a), (c), and (e) show $\vert E_{\rho 1}(r)\vert$, $\vert E_{\theta
    1}(r)\vert $, and $\vert E_{z1}(r)\vert $; (b), (d), and (f) show
  $\log_{10}$ of the estimated pointwise absolute error.}
\label{fig:n110Estf}
\end{figure}

\begin{figure}[!t]
\centering 
\includegraphics[height=74mm]{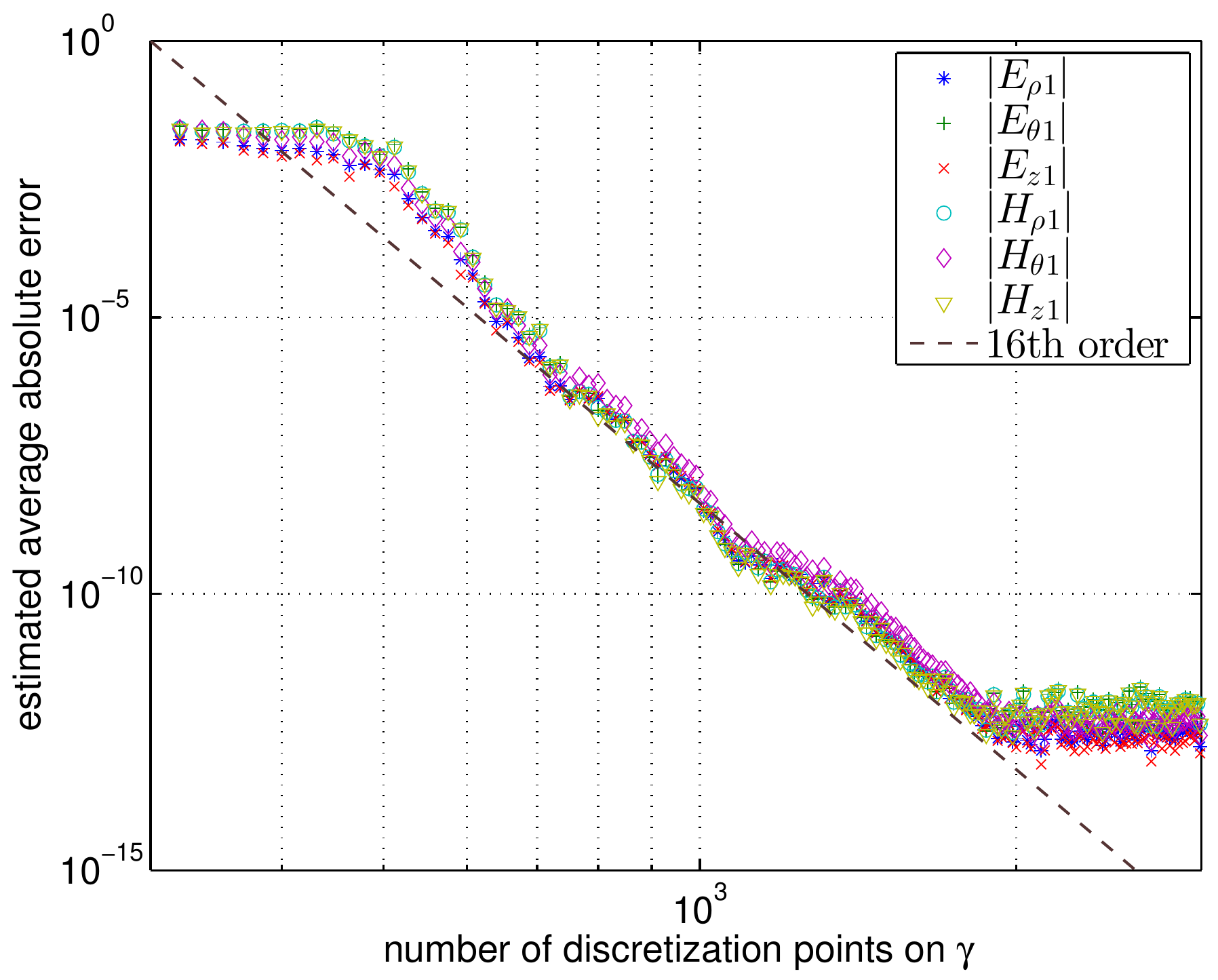}
\caption{\sf Convergence of the electric field plots shown in
  Figure~\ref{fig:n110Estf} and of the corresponding magnetic field.
  The average pointwise accuracy has converged to between 12 and 13
  digits at $1984$ discretization points on $\gamma$, corresponding to
  about 26 points per vacuum wavelength along $\gamma$.}
\label{fig:conv110}
\end{figure}
\begin{figure}[!b]
\centering 
\includegraphics[height=74mm]{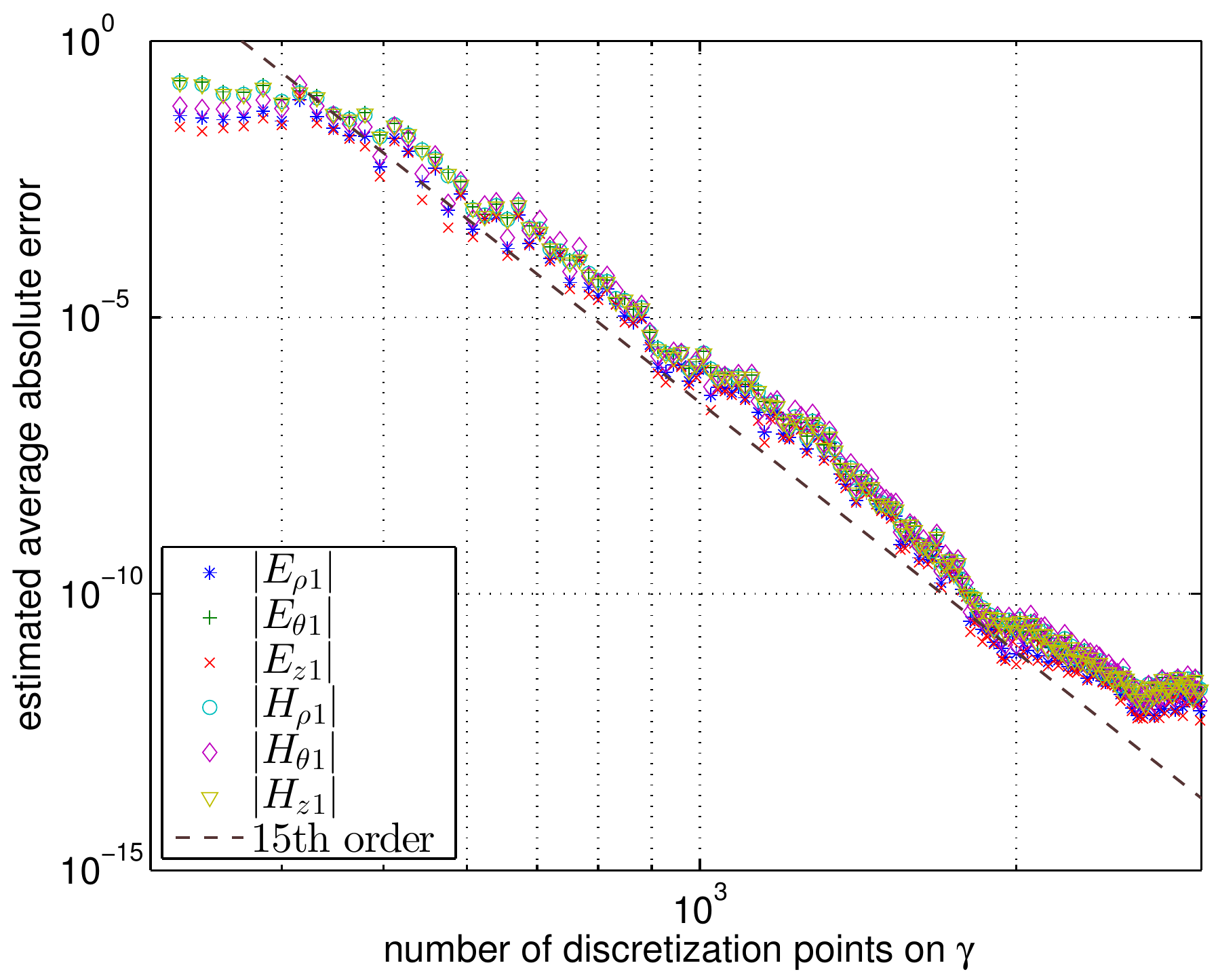}
\caption{\sf Same as in Figure~\ref{fig:conv110}, but based on the Müller
  combination~(\ref{eq:mullercomb2}). The average pointwise accuracy
  has converged to almost 12 digits at $2640$ discretization points on
  $\gamma$, corresponding to about 36 points per vacuum wavelength
  along~$\gamma$.}
\label{fig:muller}
\end{figure}

\subsection{High $k$ and small $n$}
\label{sec:hiklon}

When $\Im{\rm m}\{m\}=0$, resonances with high $k$ and small $n$ have
much smaller Q-factors than fundamental modes with similar $\Re{\rm
  e}\{k\}$. We have clearly seen this in numerical test and it can
also be understood from a phenomenological description of WGMs in
terms of internal reflections. The eigenfields with high $k$ and small
$n$ vary rapidly both outside and, in particular, inside $A$ and the
problem is harder to resolve.

Figure~\ref{fig:n110Estf} shows an example for the object in
Figure~\ref{fig:geometry} with $n=1$ and $m=1.5$. The converged
eigenwavenumber $k=110.041232211051-0.404177078290{\rm i}$ corresponds
to a generalized object diameter of about 45.9 vacuum wavelengths. The
large value of $\Im{\rm m}\{k\}$ makes the exponential growth of the
eigenfields visible already in the object's immediate vicinity.

Figure~\ref{fig:conv110} confirms that our solver exhibits 16th order
convergence and is stable under uniform overresolution. The average
pointwise accuracy in the field plots saturates at 12--13 digits,
which compares favorably with the most accurate results we have found
in the literature for general transmission problems involving axially
symmetric objects of non-trivial shapes~\cite{Liu16}.

We have done the same convergence study for our Fourier Nyström scheme
applied to \eqref{eq:mullercomb2}. The result is given in
Figure~\ref{fig:muller}. The convergence order is 15 for
\eqref{eq:mullercomb2} compared to 16 for our combination
\eqref{eq:eigenQF}. The Müller combination \eqref{eq:mullercomb2}
needs 36 discretization points per vacuum wavelength along $\gamma$
for saturated convergence, compared to 26 for \eqref{eq:eigenQF}. The
relative error of the largest absolute value of the evaluated field
components in Figure \ref{fig:n110Estf} is $1.3\cdot 10^{-12}$ for
\eqref{eq:mullercomb2} and $3.7\cdot 10^{-13}$ for \eqref{eq:eigenQF}.
The construction of the system matrix took 680 seconds for
\eqref{eq:mullercomb2} and 350 seconds for \eqref{eq:eigenQF}. It took
45 seconds to find the solution to \eqref{eq:mullercomb2}, compared to
70 seconds for \eqref{eq:eigenQF}. It took on average 0.072 seconds to
evaluate the six field components at a point $r$ for
\eqref{eq:mullercomb2} and 0.042 seconds for \eqref{eq:eigenQF}, with
\eqref{eq:eigenE1}--\eqref{eq:eigenH2}.

\begin{figure}
\centering 
\includegraphics[height=80mm]{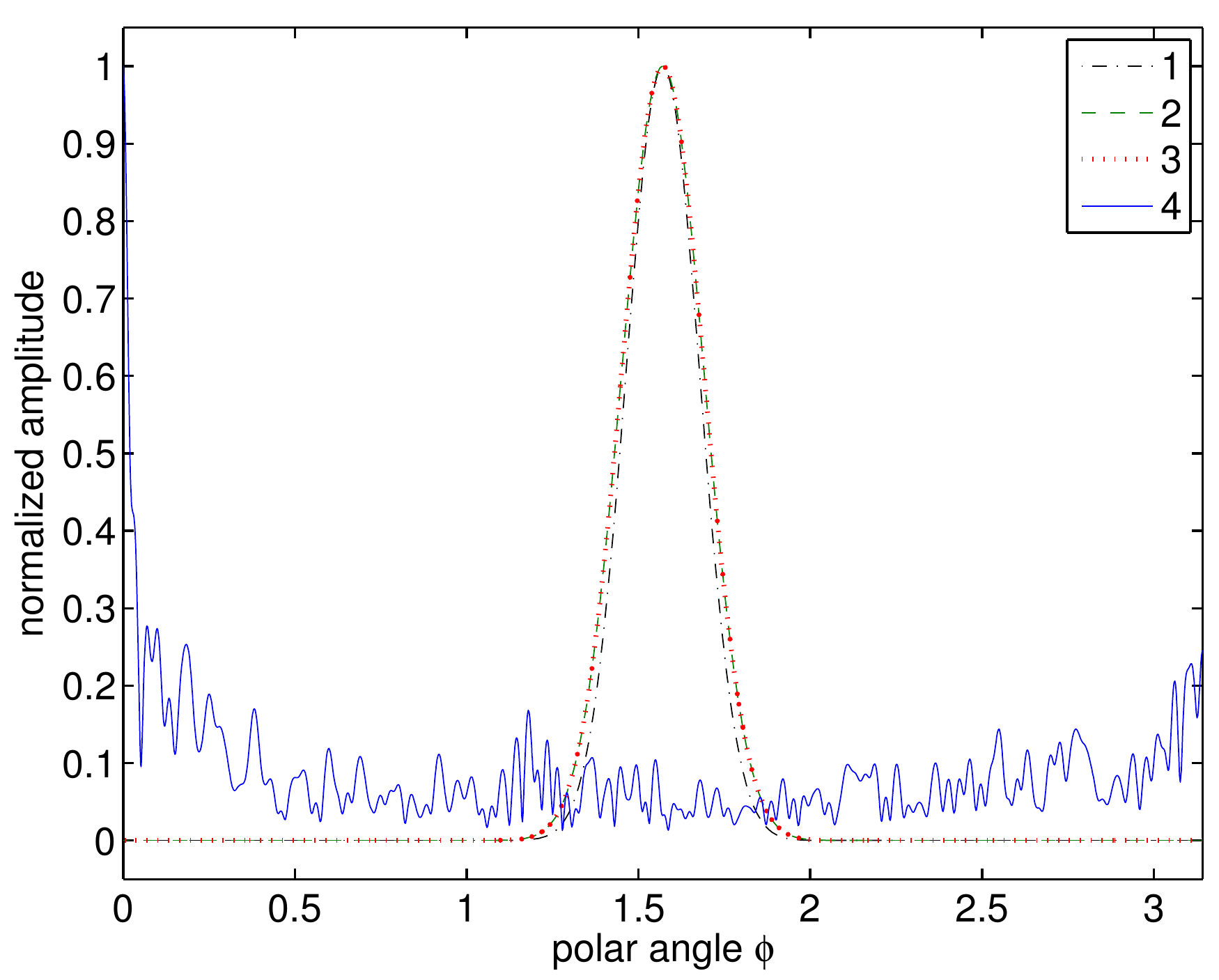}
\caption{\sf Far-field patterns: curve 1 is for the WGM of the unit
  sphere of Section~\ref{sec:unit}; curves 2 and 3, which are almost
  identical, are for the WGMs with lossless and lossy object materials
  of Section~\ref{sec:LvsL}; curve 4 is for the $n=1$ mode of
  Section~\ref{sec:hiklon}.}
\label{fig:pattern}
\end{figure}

\subsection{Far-field patterns}

Far-field patterns of resonant modes are defined by
\eqref{eq:farfield1} and \eqref{eq:farfield2}. A necessary condition
for their meaningful evaluation when $\Im{\rm m}\{m\}=0$ is that
$\Im{\rm m}\{k\}$ is known with a relative accuracy better than one
per cent. This means that if ${\rm Q}_{\rm rad}$, which for $\Im{\rm
  m}\{m\}=0$ is equal to ${\rm Q}$, is on the order of $10^p$, then
$k$ needs to be resolved with at least $p+2$ digits. This condition,
coupling the magnitude of ${\rm Q}_{\rm rad}$ to the precision
required in $k$, seems to hold also when $\Im{\rm m}\{m\}\ne 0$. The
requirement of $p+2$ accurate digits in $k$ is met in all our
examples, except for that of the high wavenumber WGM in
Section~\ref{sec:hiWGM}.

Figure~\ref{fig:pattern} shows that the far-field patterns of the WGMs
are smooth and resemble each other. Their radiated fields peak at the
equator, $\phi=\pi/2$. The variation in the pattern of the $n=1$ mode
of Section~\ref{sec:hiklon} is rapid, as expected.

\section{Conclusions}
\label{sec:concl}
Our solver,
for the determination of resonant modes of axially symmetric
dielectric objects, uses  integral equations, related to  the  Müller formulation, and charge integral equations. This, in
combination with a high-order convergent
discretization, allows for exceptionally accurate
results and excludes the possibility of finding spurious solutions.
Moreover, the solver extends the admissible size of objects for which
high accuracy can be obtained, based on the full vectorial Maxwell
equations, into the regime where asymptotic methods for WGMs are
applicable.
We stress the following capabilities of our solver up to such object
sizes:
\begin{itemize}
\item The evaluation of the entire spectrum and all eigenfields with
  $\Re{\rm e}\{k\}$ in a given interval. This includes the
  computationally difficult resonances with small $n$ and high $k$. 
\item The evaluation of eigenfields at any point in space. This
  includes slowly-evanescent and radiated fields.
\item The evaluation of far-field patterns of WGMs with radiative
  Q-factors up to $10^{13}$.
\item Geometric flexibility. While high-order surface information is a
  prerequisite, non-smooth boundaries can be treated.
\end{itemize}
These capabilities open up for new studies related to the coupling of
electromagnetic waves into WGMs, to finding new object shapes for WGMs and resonances
in objects with non-linear as well as active materials.
They also make the solver ideal for benchmarking.

\section*{Acknowledgement}

\noindent
This work was supported by the Swedish Research Council under contract
621-2014-5159.

\renewcommand{\theequation}{A.\arabic{equation}}
\setcounter{equation}{0}

\section*{Appendix A. The  operators $K_{24n}$, $K_{25n}$, and $K_{26n}$}
\label{sec:explicit}

The modal operators $K_{in}$, $i=24,25,26$, are most easily defined in
terms of~(\ref{eq:GF}) and~(\ref{eq:modalop}) and the kernels
\begin{equation}
K_i(\vec r,\vec r')=D_i(\vec r,\vec r')
(1-{\rm i}k\vert\vec r-\vec r'\vert)e^{{\rm i}k\vert\vec r-\vec r'\vert}\,,
\, i=24,25,26\,,
\end{equation}
with static factors
\begin{align}
D_{24}(\vec r,\vec r')&=\frac{\vec\tau\cdot(\vec r-\vec r')}
{4\pi\vert\vec r-\vec r'\vert^3}\,,\\
D_{25}(\vec r,\vec r')&={\rm i}
\frac{\left(\nu_\rho\nu'\cdot r'+\nu'_z\tau\cdot r\right)\sin(\theta-\theta')}
{4\pi\vert\vec r-\vec r'\vert^3}\,,\\
D_{26}(\vec r,\vec r')&=\frac{\nu_z\rho'-\left(\tau\cdot r+\nu_\rho z'\right)
\cos(\theta-\theta')}
{4\pi\vert\vec r-\vec r'\vert^3}\,.
\end{align}
and, with notation as in~\cite[Appendix~A]{HelsKarl16},
corresponding Fourier coefficients 
\begin{align}
D_{24n}(r,r')&=-\eta\left[d(\tau)\mathfrak{R}_n(\chi)
-\frac{\nu_z}{\rho}\mathfrak{P}_n(\chi)\right]\,,
\label{eq:D24n}\\
D_{25n}(r,r')&=\eta
\frac{\left(\nu_\rho\nu'\cdot r'+\nu'_z\tau\cdot r\right)}
{\rho\rho'}n\mathfrak{Q}_{n-\frac{1}{2}}(\chi)\,,
\label{eq:D25n}\\
D_{26n}(r,r')&=\eta\left[
d(\tau)\mathfrak{R}_n(\chi)
+\frac{\left(\tau\cdot r+\nu_\rho z'\right)}{\rho\rho'}
\mathfrak{P}_n(\chi)\right]\,.
\label{eq:D26n}
\end{align}
The reason for including~(\ref{eq:D24n})-(\ref{eq:D26n}) in this
exposition is that these expression are used in the convolutions in
our numerical scheme, see Section~\ref{sec:over}.

\begin{small}

\end{small} 
\end{document}